\documentclass[10pt]{article}
\usepackage[normalem]{ulem}
\usepackage[utf8]{inputenc}
\usepackage[english]{babel}
\usepackage[toc,page]{appendix}

\usepackage{amsmath,amsfonts,amssymb,amsthm}
\usepackage{graphicx}
\usepackage{float}
\usepackage{hyperref}
\usepackage{dsfont}
\usepackage{subcaption}
\usepackage[format=plain,font=small,textfont=it]{caption}
\usepackage[dvipsnames]{xcolor}
\usepackage{multirow} 
\usepackage{bookmark}
\usepackage{mathrsfs}
\usepackage{physics}

\usepackage[sorting=none]{biblatex}
\graphicspath{{img/}}

\usepackage[colorinlistoftodos,prependcaption,textsize=tiny]{todonotes}

\newcommand{\CC}{\mathbb{C}}
\newcommand{\RR}{\mathbb{R}}

\newcommand{\conte}{\mathcal{C}_\varepsilon}

\newcommand{\nn}{\nonumber}

\newcommand{\FJM}[2]{\left(\begin{matrix}
#1\\
#2
\end{matrix}\right)}

\newcommand{\spinF}{{$Spin(4)$}}
\newcommand{\sltc}{{$SL(2,\mathbb{C})$}}
\newcommand{\SUT}{{$SU(2)$}}

\begin{document}


\title{A Wick rotation for EPRL spin foam models}

\author{\Large{Pietro Don\`a${}^{a}$\footnote{dona@cpt.univ-mrs.fr}, \ } \Large{Francesco Gozzini${}^{a}$\footnote{gozzini@cpt.univ-mrs.fr}, \ } \Large{Alessandro Nicotra${}^{b}$\footnote{alessandro.nicotra5@studio.unibo.it} \ }
\smallskip \\ 
\small{\textit{a CPT, Aix-Marseille\,Universit\'e, Universit\'e\,de\,Toulon, CNRS, 13288 Marseille, France}}\\
\small{\textit{b DIFA, Alma Mater Studiorum, Universit\`a di Bologna, Italy}}
}

\date{}

\maketitle

\begin{abstract}
\noindent We show that the Euclidean and Lorentzian EPRL vertex amplitudes of covariant Loop Quantum Gravity are related through a ``Wick rotation'' of the real Immirzi parameter to purely imaginary values. Our result follows from the simultaneous analytic continuation of the algebras, group elements and unitary irreducible representations of the gauge groups \spinF{} and \sltc{}, applied to the decomposition of the two models in terms of \SUT{} invariants and booster functions.
\end{abstract}


\section{Introduction}
\label{sec:intro}

Spin foam theory attempts to define the Loop Quantum Gravity dynamics with a regularized, background-independent, and Lorentz covariant quantum gravity path integral on a fixed triangulation.  
The most popular spin foam model is the EPRL-FK model defined in \cite{Engle:2007wy, Freidel:2007py} (see \cite{Perez:2012wv} or \cite{Rovelli2015} for more pedagogical reviews). Both models naturally have as boundary states the spin network states from canonical Loop Quantum Gravity, which is why spin foam theory is considered ``covariant Loop Quantum Gravity'' \cite{Rovelli:2010wq}. 
The theory assigns transition amplitudes to spin network states living on the boundary of four-dimensional triangulations. A version of the model exists for spacetimes with both Euclidean and Lorentzian signatures.

\medskip

The two models differ by their gauge group structures: \spinF{} (the double covering of SO(4)) for the Euclidean signature and \sltc{} (the double covering of $SO_+(3,1)$) for the Lorentzian one. The models are built upon unitary irreducible representations of the groups (the principal series ones in the Lorentzian case). The linear simplicity constraints, responsible for reducing the BF topological theory to General Relativity, are implemented weakly as a restriction on the representation labels.

Performing calculations with the Euclidean model is much simpler. The gauge group is compact, and it has finite-dimensional unitary irreducible
representations. Moreover, $Spin(4) \simeq SU(2)\times SU(2)$, so its irreducible representations can be written as the tensor product of \SUT{} ones. Those are very well studied, and the amplitude is the contraction of \SUT{} invariants. 
The computations within the Lorentzian EPRL model are notably more complicated than in the Euclidean one. The group is non-compact, therefore its unitary representations are infinite-dimensional and much less studied. It is not surprising that the Euclidean model is the preferred first choice to perform complex calculations. Given the similarities between the two models, it has been so far assumed, as a strong hypothesis, that the results obtained in the simpler Euclidean model also hold for the Lorentzian one. The alternative is to repeat the specific calculation from scratch in the Lorentzian setting. 

\medskip

This work provides a map between the constrained irreducible representations and the spin foam vertex amplitudes of the two models. The map is realized through the ``rotation'', or analytic continuation, of the real Immirzi parameter $\gamma$ to purely imaginary values. We show that the Euclidean vertex amplitude can be analytically continued to the Lorentzian amplitude, up to a multiplicative factor, sending $\gamma \to i\gamma$. Conversely, the Lorentzian amplitude continues to the Euclidean one with the inverse map $i\gamma \to \gamma$. We refer to this map as ``Wick rotation'' in analogy to the signature changing transformation in Quantum Field Theory. We remark that our transformation is a map between the internal gauge groups \spinF{} and \sltc{} or equivalently a rotation between Euclidean and Lorentzian Ashtekar-Barbero variables \cite{Immirzi:1996di,Ashtekar:1995qw, barberoWick}, not on the spacetime manifold. This one has to be appropriately reconstructed in the semiclassical limit.

\medskip

The paper is organized as follows. In Section \ref{sec:math}, we review the \sltc{} and \spinF{} algebras, group representations, useful decompositions, and the matrix elements of unitary irreducible representations. This is a well-known subject in the LQG community. However, the presentation for the two groups is traditionally very different. We take this occasion to fix our notations while treating the two groups along similar lines. In particular, we introduce the canonical basis for \spinF{}.  In Section \ref{sec:models}, we review the definition and implementation of the $Y_\gamma$ map, the main ingredient in the construction of the EPRL model. We also review the decomposition of the vertex amplitude in terms of booster functions. The EPRL models with different signatures differ only by the definition of their booster functions. The original contribution of this work is in Section \ref{sec:map}. From a map between the algebras of \sltc{} and \spinF{} we derive a map between unitary irreducible representations and group elements. We show that the matrix elements of the $(\rho,k)$ representation of \sltc{}, with $\rho\in\RR$ and $k\in \mathbb{Z}/2$, can be obtained by analytic continuation from the matrix elements of the $(p,k)$ representation of \spinF{}, with $p\in\mathbb{Z}/2$ and $k\in \mathbb{Z}/2$. The analytic continuation is performed simultaneously on the representation labels $(p,k) \leftrightarrow (i\rho,k)$ and on the group elements. We define the analytic continuation of the group elements using the Cartan decomposition of both groups. 
Using these maps between group elements and representations, we show that the Euclidean and Lorentzian booster functions are related through analytic continuation when they are defined as line integrals along a particular complex contour. We conclude by applying this analytic continuation to the constrained representations of the vertex amplitudes with both signatures.

\section{Mathematical preliminaries}
\label{sec:math}


\subsection{Representations of \sltc{}}
\label{sec:math-SL2C}

The algebra of \sltc{} is generated by $L_i$, the generators of the spatial rotation subgroup, and $K_i$, the generators of the corresponding boosts. They satisfy the commutation relations
\begin{equation}
\label{eq:sl2calgebra}
\left[L_{i},L_{j}\right]=i\epsilon_{ijk}L_{k} \ ,\qquad
\left[L_{i},K_{j}\right]=i\epsilon_{ijk}K_{k} \ ,\qquad
\left[K_{i},K_{j}\right]=-i\epsilon_{ijk}L_{k} \ .
\end{equation}
The two Casimir operators are $K^2 - L^2$ and $\vec{K}\cdot\vec{L}$. The unitary irreducible representations in the principal series are labeled by the couple $(\rho,k)$ where $\rho$ is a real number and $k$ is a half-integer. The Casimirs in this representations take the values 
\begin{equation}
\label{eq:CasimirSL2C}
\left(K^2 - L^2\right)\ket{\rho,k}= (\rho^2-k^2+1)\ket{\rho,k} \ , \qquad \vec{K}\cdot\vec{L}\ket{\rho,k}=\rho k \ket{\rho,k} \ .
\end{equation}
The representation $(-\rho,-k)$ is unitarily equivalent to the representation $(\rho,k)$ \cite{Ruhl1970}. This property is manifest in the expression for the matrix elements we derive in Section \ref{sec:map-matrixel}. Following the literature on the topic \cite{Ruhl1970} we restrict ourselves to only positive values of $\rho$ and $k$. 

\medskip

The group \sltc{} is non-compact so the generic unitary representation $(\rho,k)$ is infinite dimensional. The Hilbert space $\mathcal{H}^{(\rho,k)}$ of the representation $(\rho,k)$ decomposes in an infinite number of \SUT{} representations with different values $j$ of $L^2$
\begin{equation}
\label{eq:SU2TowerSL2C}
\mathcal{H}^{(\rho,k)} = \bigoplus_{j=k}^\infty \mathcal{H}^{j} \ .
\end{equation}
The canonical basis of $(\rho,k)$, given by $\ket{\rho,k; j m}$ with $j\geq k$ and $m = -j,\ldots,j$, diagonalizes $L^2$ and $L_3$
\begin{equation}
L^2\ket{\rho,k;j,m}= j(j+1)\ket{\rho,k;j,m} \ , \qquad L_3\ket{\rho,k;j,m}= m\ket{\rho,k;j,m} \ .
\end{equation}
The canonical basis plays a central role in the construction of the EPRL model. 

\medskip

A useful decomposition of the group \sltc is given by the map
\begin{equation}
\begin{split}
\label{eq:CartanSL2C}
  SU(2) \times A^+ \times SU(2) &\longrightarrow  SL(2,\CC) \\
  (u,e^{\frac{r}{2} \sigma_3},v)   &\longrightarrow ue^{\frac{r}{2} \sigma_3} v^\dagger
   \ .
\end{split}
\end{equation}
where $A^+$ is the diagonal subgroup
\begin{equation}
  A^+: e^{\frac{r}{2} \sigma_3} = \begin{pmatrix}
    e^{r/2} & 0 \\
    0 & e^{-r/2}
  \end{pmatrix}, \quad r \geq 0 \ .
\end{equation}
This is usually called \emph{Cartan decomposition} in the physics literature \cite{Ruhl1970,Speziale:2016axj} \footnote{While in the mathematical literature this is referred to either as the \emph{KAK decomposition} or as the \emph{polar decomposition}.}. The Haar measure with respect to this decomposition is \cite{Ruhl1970,Speziale:2016axj} 
\begin{equation}
\label{eq:HaarSL2C}
 \dd\mu_{SL(2,\CC)} =  \frac{1}{\pi} \sinh^2 r \ \dd r \ \dd u\  \dd v = \dd \mu(r)\ \dd u\  \dd v \ .
\end{equation}
The normalization factor $\frac{1}{\pi}$ we use in this paper differs from the one used in the literature by a factor $4$. This choice allows us to write our formulas more cleanly. The matrix elements in the canonical basis of a group element $g$ in this decomposition also decompose accordingly
\begin{equation}
D_{jmln}^{(\rho,k)}(g) \equiv \bra{\rho,k;j,m} g \ket{\rho,k;l,n} =  D_{jmln}^{(\rho,k)}(ue^{\frac{r}{2} \sigma_3}v^\dagger) = \sum_{a,a'} D_{ma}^{j}(u)D_{jala'}^{(\rho,k)}(e^{\frac{r}{2} \sigma_3}) D_{a'n}^{l}(v^\dagger) \ .
\end{equation}
In the expression above we used that the \SUT{} subgroup is generated by $\vec{L}$ and its matrix elements are given by \SUT{} Wigner matrices (see Appendix~\ref{app:SU2})
\begin{equation}
\bra{\rho,k;j,m} u \ket{\rho,k;l,n} = \delta_{jl} D^j_{mn}(u) \ .
\end{equation}
Moreover, $e^{\frac{r}{2} \sigma_3}$ is diagonal, therefore $D_{jmln}^{(\rho,k)}(e^{r\sigma_3})= \delta_{aa'} D_{jala}^{(\rho,k)}(e^{r\sigma_3}) \equiv \delta_{aa'} d_{jla}^{(\rho,k)}(r)$ where $d_{jla}^{(\rho,k)}$ are the reduced matrix elements of \sltc. Summarizing
\begin{equation}
\label{eq:CartanDSL2C}
D_{jmln}^{(\rho,k)}(g) = \sum_{a} D_{ma}^{j}(u)\ d_{jla}^{(\rho,k)}(r)\ D_{an}^{j}(v^\dagger) \ .
\end{equation}

\medskip

The expression for $d_{jlm}^{(\rho,k)}(r)$ was given in \cite{Ruhl1970,Speziale:2016axj,Dao:1967ri,Rashid:1979xv,Basu:1977ii}
\begin{equation}\label{eq:dSL2C}
\begin{split}
d_{jlm}^{(\rho,k)}(r)&=(-1)^{j-l}\sqrt{\frac{\left(i\rho-j-1\right)!\left(j+i\rho\right)!}{\left(i\rho-l-1\right)!\left(l+i\rho\right)!}}\frac{\sqrt{(2j+1)(2l+1)}}{(j+l+1)!}e^{(i\rho-k-m-1)r}\\
&\sqrt{(j+k)!(j-k)!(j+m)!(j-m)!(l+k)!(l-k)!(l+m)!(l-m)!}\\
&\sum_{s,t}(-1)^{s+t}e^{-2tr}\frac{(k+s+m+t)!(j+l-k-m-s-t)!}{t!s!(j-k-s)!(j-m-s)!(k+m+s)!(l-k-t)!(l-m-t)!(k+m+t)!}\\
&{}_{2}F_{1}\left[\{l-i\rho+1,k+m+s+t+1\},\{j+l+2\};1-e^{-2r}\right]
\end{split}
\end{equation}
where ${}_{2}F_{1}$ is the Gauss hypergeometric function, and the factorials of complex numbers in this formula and the rest of this paper have to be intended as Gamma functions using $x! = \Gamma(x+1)$.


\subsection{Representations of \spinF{}}
\label{sec:math-Spin4}


The algebra of $Spin(4)=SU(2)\times SU(2)$ is generated by two commuting $SU(2)$ algebras
\begin{equation}
\left[J_L^i,J_L^j\right]=i\epsilon_{ijk}J_L^k \ ,\qquad
\left[J_R^i,J_R^j\right]=i\epsilon_{ijk}J_R^k \ ,\qquad
\left[J_L^i,J_R^j\right]=0 \ .
\end{equation}
The two Casimir operators are $J_L^2$ and $J_R^2$ and the unitary irreducible representations are labeled by the couples $(j_L,j_R)$ of half-integers such that%

\begin{equation}
\label{eq:StandardBasis}
J_L^2\ket{j_L,j_R}= j_L(j_L+1)\ket{j_L,j_R} \ , \qquad J_R^2\ket{j_L,j_R}= j_R(j_R+1)\ket{j_L,j_R} \ .
\end{equation}
The group \spinF{} is compact and the unitary representation $(j_L,j_R)$ has dimension $(2j_L+1)(2j_R+1)$. The algebra of \spinF{} is isomorphic to the algebra of $SO(4)$. It is convenient to parametrize the algebra in terms of $L_i$, the generators of the spatial rotation subgroup, and $A_i$, the generators of time rotations or (Euclidean) boosts, as we will call them with a slight abuse of language. Defining the rotations and boost generators as
\begin{equation}
\vec{L}=\vec{J}_L + \vec{J}_R \ , \hspace{1cm} \text{and}  \hspace{1cm} \vec{A}=\vec{J}_L - \vec{J}_R \ ,
\end{equation}
they satisfy the algebra
\begin{equation}
\label{eq:so4algebra}
\left[L_{i},L_{j}\right]=i\epsilon_{ijk}L_{k} \ ,\qquad 
\left[L_{i},A_{j}\right]=i\epsilon_{ijk}A_{k} \ , \qquad
\left[A_{i},A_{j}\right]=i\epsilon_{ijk}L_{k} \ .
\end{equation}
We can sum and subtract the Casimirs \eqref{eq:StandardBasis} to obtain an equivalent set of two \spinF{} invariant operators
\begin{equation}
A^2+L^2=2\left(J_L^2+J_R^2\right)\ , \qquad \vec{L}\cdot\vec{A}=J_L^2-J_R^2 \ .
\end{equation}
We parametrize the representation $(j_L,j_R)$ in terms of two other half integer quantum numbers $p\equiv j_L+j_R+1$ and $k\equiv j_L-j_R$ \cite{Biedenharn}\footnote{We use a slightly different definition from \cite{Biedenharn}, in which $p\equiv j_L+j_R$.}. In this work, without any loss of generality we will assume that $j_L\geq j_R$ such that $p>k\geq 0$. In this representation, the Casimirs assume the values
\begin{equation}
\left(A^2+L^2\right)\ket{p,k}=(p^2+k^2-1)\ket{p,k} \ , \qquad \vec{L}\cdot\vec{A}\ket{p,k} = p k \ket{p,k} \ .
\end{equation}
The representation space $\mathcal{H}^{(p,k)}$ decomposes in \SUT{} representations with different values of $L^2=(\vec{J}_L + \vec{J}_R)^2$ given by the usual sum of $L$(eft) and $R$(ight) angular momentum
\begin{equation}
\mathcal{H}^{(p,k)} = \bigoplus_{j=k}^{p-1} \mathcal{H}^{j} \ .
\end{equation}
The canonical basis of $(p,k)$, given by $\ket{p,k; j m}$ with $p-1\geq j\geq k$ and $m = -j,\ldots,j$, diagonalizes $L^2$ and $L_3$
\begin{equation}
L^2\ket{\rho,k;j,m}= j(j+1)\ket{\rho,k;j,m} \ , \qquad L_3\ket{\rho,k;j,m}= m\ket{\rho,k;j,m} \ .
\end{equation}
Note that, in the spin foam literature, the Euclidean EPRL model is often formulated using the standard basis that diagonalizes $J_{L3}$, $J_{R3}$ while $J_L^2$ and $J_R^2$ are taken as Casimirs. To highlight the similarities between the models with the two signatures, we formulate the Euclidean EPRL model in the canonical basis instead. We stress that our construction uses a different language from the traditional formulation, but it is entirely equivalent to it. 

\medskip

We want to find a decomposition analogue to the Cartan decomposition \eqref{eq:CartanSL2C} for the group \spinF{}. Inspired by the Euler parametrization of the rotation group, we look at the action
\begin{equation}
(g_L,g_R) \cdot h = g_L h g_R^\dagger 
\end{equation}
of $Spin(4) \simeq SU(2) \times SU(2)$ on the group $\mathbb{H}$ of quaternions. This action realizes the double-covering $Spin(4) \to SO(4)$ and in particular it is transitive on $SU(2) \simeq S^3$ in $\mathbb{H}$. The diagonal subgroup $(a,a) = D \simeq SU(2) \subset Spin(4)$ is the stabilizer of the identity. We parametrize an arbitrary element $(g_L,g_R) \in Spin(4)$ using two copies of this subgroup. We define the map 
\begin{equation}
\begin{split}
  SU(2) \times T \times SU(2) &\longrightarrow Spin(4) \\
(u,e^{ -i \frac{t}{2} \sigma_3}, v) &\longrightarrow  (ue^{ -i \frac{t}{2} \sigma_3}v^\dagger, u e^{i \frac{t}{2} \sigma_3}v^\dagger)
\end{split}
\end{equation}
where $T = \{\exp(-i \frac{t}{2} \sigma_3)\ |\ t \in [0\ 4\pi)\}$ is the torus subgroup of \SUT{}. We show that the map is surjective. Let $(g_L,g_R)$ be a generic element of $Spin(4)$. The equations
\begin{align*}
  g_L &= ue^{ -i \frac{t}{2} \sigma_3}v^\dagger \\
  g_R &= ue^{ i \frac{t}{2} \sigma_3}v^\dagger
\end{align*}
imply
\begin{align*}
  g_Lg_R^\dagger &= ue^{ -i t \sigma_3}u^\dagger \\
  g_R^\dagger g_L &= v e^{ -i t \sigma_3}v^\dagger.
\end{align*}
The elements $g_Lg_R^\dagger$ and $g_R^\dagger g_L$ are conjugate, and every element of $SU(2)$ is conjugate to a diagonal matrix of the form $\exp (i t \sigma_3)$. Hence we can solve the last equations for $u,v$. Notice, importantly, that it is enough to require $t \in [0\ 2\pi)$ to get a unique solution. Therefore the proper \textit{Cartan decomposition} for \spinF{} is
\begin{equation}
\label{eq:CartanSpin4}
\begin{split}
  SU(2) \times T^+ \times SU(2) &\longrightarrow Spin(4) \\
(u,e^{ -i \frac{t}{2} \sigma_3}, v) &\longrightarrow  (ue^{ -i \frac{t}{2} \sigma_3}v^\dagger, u e^{i \frac{t}{2} \sigma_3}v^\dagger)
\end{split}
\end{equation}
where $T^+ = \{\exp(-i \frac{t}{2} \sigma_3)\ |\ t \in [0\ 2\pi)\}$. In the following we will also use the notation $E^+ = \{ (g, g^\dagger)\ |\ g \in T^+ \}$.

The Haar measure with respect to the decomposition described in \eqref{eq:CartanSpin4} is left and right $(D \simeq SU(2))$-invariant. It is easy to show that the measure must be of the form
\begin{equation}
  \dd\mu_{Spin(4)} =  N f(t) \ \dd t \ \dd u\  \dd v \ ,
\end{equation}
with $\dd u, \dd v$ the usual Haar measure of \SUT{} and $N$ a normalization constant that we fix to obtain total unit volume. A Jacobian computation similar to the \sltc{} case shows that $f(t) = \sin^2 t$, and normalizing we get the Haar measure
\begin{equation}
\label{eq:HaarSpin4}
  \dd\mu_{Spin(4)} =  \frac{1}{\pi} \sin^2t \ \dd t \ \dd u\  \dd v =   \dd \mu(t) \ \dd u\  \dd v \ .
\end{equation}

\medskip

The matrix elements of $g=(g_L,g_R)$ in the standard basis are given by the tensor product of two $SU(2)$ Wigner matrices
\begin{align}
\label{eq:Dstandard}
D_{m_Lm_Rn_L n_R}^{(j_L,j_R)}(g) \equiv& \bra{j_L,m_L;j_R,m_R} g \ket{j_L,n_L;j_R,n_R} =  \notag \\
=&\bra{j_L,m_L} g_L \ket{j_L,n_L} \bra{j_R,m_R} g_R \ket{j_R,n_R} = D^{j_L}_{m_Ln_L}(g_L)D^{j_R}_{m_Rn_R}(g_R)  \ .
\end{align}
This is the main reason why the $\ket{j_L, j_R}$ basis is traditionally preferred for the construction of the EPRL model. The Cartan decomposition for \spinF{} \eqref{eq:CartanSpin4} allows us to decompose the matrix elements in the canonical basis similarly to \sltc{}
\begin{equation}
\label{eq:CartanDSpin4}
D_{jmln}^{(p,k)}(g) = \sum_{o} D_{mo}^{j}(u)d_{jlo}^{(p,k)}(t) D_{on}^{l}(v^\dagger) \ .
\end{equation}
In the expression above we used that the $D\simeq SU(2)$ is generated by $\vec{L}$, the matrix elements of $u\in SU(2)$ are given by
\begin{equation}
\bra{p,k;j,m} u \ket{p,k;l,n} = \braket{p,k;j,m}{j_L m_L j_R m_R}  D^{j_L}_{m_Ln_L}(u)D^{j_R}_{m_Rn_R}(u) \braket{j_L n_L j_R n_R}{p,k;l,n} = \delta_{jl} D^j_{mn}(u) \ .
\end{equation}
The reduced matrix elements $d_{jlo}^{(p,k)}(t)$ are defined in terms of SU(2) Clebsch-Gordan coefficients intertwining between the $(j_L,j_R)$ and $j$ or $l$ $SU(2)$ representation \cite{Biedenharn,Lorente}
\begin{align}
\label{eq:dSpin4}
d_{jlm}^{(p,k)}(t)&=D_{jmlm}^{(p,k)}(e^{i\frac{t}{2}\sigma_3},e^{-i\frac{t}{2}\sigma_3})=\sum_{m_L,m_R} \braket{j_L,m_L,j_R,m_R}{jm} e^{it(m_L-m_R)}\braket{j_L,m_L,j_R,m_R}{lm} \\
&= \sum_{n=0}^{2p-2q-1}e^{it(p-q-1+m-2n)} \braket{\frac{p+q-1}{2},\frac{p-q-1}{2}+m-n,\frac{p-q-1}{2},n-\frac{p-q-1}{2}}{jm} \nn \\
&\hspace{4.2cm}\braket{\frac{p+q-1}{2},\frac{p-q-1}{2}+m-n,\frac{p-q-1}{2},n-\frac{p-q-1}{2}}{lm} \ .\nn
\end{align}


\section{The Lorentzian and the Euclidean EPRL model}
\label{sec:models}


The starting point for constructing the EPRL model is the spin foam quantization of a topological BF theory leading to a well-defined state sum model. The classical simplicity constraints reduces the topological BF theory to gravity. The path integral for quantum gravity is obtained implementing the quantum simplicity constraints on the BF partition function. The simplicity constraints are expressed by an equation involving non-commuting operators and cannot be imposed strongly. The solution is to implement them weakly using master constraints techniques involving Casimir operators. The result is a restriction on the unitary representations that contribute to the state sum model.

We will very briefly review the implementation of the linear simplicity constraints via the $Y_\gamma$ map in both the Lorentzian and Euclidean EPRL model. For an exhaustive discussion, we refer to the original paper \cite{Engle:2007wy} or the reviews \cite{Perez:2012wv, Rovelli2015}. The goal of the following sections is to fix the notation used in this work and to show the similarities in the implementation of the $Y_\gamma$ if we use the canonical basis for the groups with both signatures. 


\subsection{The Lorentzian $Y_\gamma$ map}


The linear simplicity constraints in the Lorentzian model impose a linear dependence between the rotation and boost generators of the \sltc{} algebra \eqref{eq:sl2calgebra}
\begin{equation}
\label{eq:linearSconstraints}
\vec{K} = \gamma \vec{L} \ .
\end{equation}
However, as rotation and boost generators do not commute, such equation cannot be imposed strongly at the quantum level. We apply it weakly using two master constraints by imposing strongly two commuting quadratic operators derived from \eqref{eq:linearSconstraints}. We square \eqref{eq:linearSconstraints} and project it onto $\vec{L}$ to obtain two constraints written in terms of the Casimirs of \sltc{} \eqref{eq:CasimirSL2C}
\begin{equation}
\label{eq:constrCasimirSL2C}
K^2-L^2=(\gamma^2-1)L^2 \ , \qquad \vec{L}\cdot\vec{K}=\gamma L^2 \ .
\end{equation}
On the canonical basis of the unitary irreducible representation of \sltc{} in the principal series the constraints \eqref{eq:constrCasimirSL2C} translate to an equation for the representation labels
\begin{equation}
\rho^2-k^2+1=(\gamma^2-1)j(j+1) \ , \qquad \rho k=\gamma j(j+1) \ .
\end{equation}
For large quantum numbers this equation is solved by
\begin{equation}
\label{eq:gammasimpleSL2C}
\rho=\gamma j \  \qquad \text{and} \qquad k=j \ .
\end{equation}
The constraints \eqref{eq:constrCasimirSL2C} select special representations of \sltc{} (with representation labels proportional by a factor $\gamma$) and project to the lowest \SUT{} subgroup of \eqref{eq:SU2TowerSL2C}. Equations \eqref{eq:gammasimpleSL2C} define a map from the \SUT{} representation of spin $j$ to a subspace of the \sltc{} representation $(\gamma j,j)$
\begin{equation}
Y_\gamma:\ket{ j,m} \rightarrow \ket{\gamma j, j;  j,m } \ .
\end{equation}
Therefore, $Y_\gamma$ is a map from $SU(2)$ spin networks, the kinematical states of LQG, to \sltc{} spin networks, the states at the boundary of a spin foam vertex amplitude.


\subsection{The Euclidean $Y_\gamma$}

The linear simplicity constraints in the Euclidean model are very similar to the Lorentzian case. The generators of the Euclidean boosts $\vec{A}$ take the place of their Lorentzian counterpart $\vec{K}$
\begin{equation}
\label{eq:linearSconstraintsE}
\vec{A} = \gamma \vec{L} \ .
\end{equation}
The master constraints are
\begin{align}
L^2+A^2=(\gamma^2+1)L^2  \ , \qquad \vec{L}\cdot\vec{A}=\gamma L^2 \ ,
\end{align}
that on the canonical basis reduce to an equation between representation labels
\begin{align}
\label{eq:qnumSpin4}
p^2+k^2-1&=(\gamma^2+1)j(j+1) \ , \qquad pk=\gamma j(j+1) \ ,
\end{align}
that for large quantum numbers is solved by $p=\gamma j$
\footnote{As a side comment, since we are interested in solving \eqref{eq:qnumSpin4} for large quantum numbers, we could equivalently take $p=\gamma j +1$ and $k=j$ as solution of \eqref{eq:qnumSpin4}. The advantage is that the matrix element $D^{(\gamma j + 1, j)}_{jnjm}(g)=1$ for $j=0$ for all group elements. This allows to define a cylindrically consistent Euclidean EPRL model. Unfortunately, there are no obvious analog choices for the Lorentzian model. A possibility is to define the Lorentzian $Y_\gamma$ map as an analytic continuation of the representations $\rho = \gamma j + i$ and $k=j$. We leave this as a speculative comment.} and $k=j$. To be consistent with the convention of taking $p>k$ we restrict to the case of $\gamma > 1$. Notice that this is not a limitation. The case with $\gamma \leq 1$ can be studied by considering the case $p<k$ and have as solution of \eqref{eq:qnumSpin4} $p=j$ and $k=\gamma j$. We refrain to consider this case just for convenience, all the formulas in that case can be obtained by exchanging $p$ with $k$. 

The Euclidean $Y_\gamma$ map is defined as
\begin{equation}
\label{eq:EuclideanY}
Y_\gamma:\ket{ j,m} \rightarrow \ket{\gamma j, j;  j,m } \ ,
\end{equation}
and also in this case it provides a map between the LQG kinematical states to the states at the boundary of the spin foam vertex amplitude. If we insist that the map is valid for any \SUT{} irrep $j$ we also have an accidental quantization condition over the Immirzi parameter $\gamma$ as the labels $p$, $k$ and $j$ are half-integers \cite{Engle:2007wy,Freidel:2007py}.

At this point, the reader experienced with the presentation of the Euclidean EPRL model found in the literature could feel disoriented. Usually, the $Y_\gamma$ map in the Euclidean model is not defined on the canonical basis but on the standard one. In the standard basis the representation labels \eqref{eq:StandardBasis} are $j_L = \frac{p+k-1}{2}$ and $j_R = \frac{p-k-1}{2}$. If we consider representations with $p=\gamma j$ and $k=j$ we find the familiar restriction on the representations of \spinF{} with $\gamma>1$\footnote{If we take the solution of \eqref{eq:qnumSpin4} for large quantum number to be $p=\gamma (j +1)$ and $k=j$ instead as in \cite{Perez:2012wv}, we have to add an extra $\frac{\gamma}{2}$ to both $j_L$ and $j_R$.}
\begin{equation}
j_L = (\gamma +1) \frac{j}{2} - \frac{1}{2} \qquad \text{and} \qquad j_R = (\gamma -1) \frac{j}{2} - \frac{1}{2} \ .
\end{equation}
On the standard basis, we have the advantage of a simpler form of the matrix elements \eqref{eq:Dstandard}. The price to pay is to hide the parallelism with the Lorentzian version of the model, which becomes evident if we use the same (canonical) basis for both.


\subsection{Decomposition of the vertex amplitude in terms of booster functions}


The Cartan decomposition of the \sltc{} group elements \eqref{eq:CartanSL2C} allows us to recast the EPRL Lorentzian vertex amplitude as a superposition of \SUT{} $\{15j\}$ symbols weighted by the product of four booster functions
\begin{equation}
\label{eq:amplLor}
A^L_v(j_f,i_e)=\sum_{l_f,k_e}\left(\prod_e d_{k_e}B_4^L(j_f,l_f,i_e,k_e)\right)\{15j\}(l_f,k_e,i')
\end{equation}
This decomposition was first introduced in \cite{Speziale:2016axj} and is one of the fundamental ingredients of the numerical calculations performed within the model \cite{Dona:2017dvf,Dona:2019dkf,Gozzini:2019kui,Dona:2020tvv}.
The Lorentzian booster functions $B_4^L$ are the integral over $A^+$ of the product of four \sltc{} reduced matrix elements in the $(\gamma j_f, j_f)$ representation contracted with $4jm$-symbols \eqref{4jm} and have been extensively studied numerically \cite{Speziale:2016axj,Dona:2018nev,SLNEXT,Dona:2018pxq} and analytically \cite{Dona:2020xzv}:
\begin{equation}\label{SL2C_Boost}
B^L_4(j_a,l_a,i,k)\equiv
\sum_{m_a} \FJM{j_a}{m_a}^{(i)}
\int_0^{\infty}d\mu(r)\prod_a d^{(\gamma j_a,j_a)}_{j_al_am_a}(r)
\FJM{l_a}{m_a}^{(k)}
 \ ,
\end{equation}
where $d^{(\gamma j_a,j_a)}$ are the reduced matrix elements \eqref{eq:dSL2C} and $d\mu(r)$ is the part over $A^+$ of the \sltc{} Haar measure \eqref{eq:HaarSL2C}. 

\medskip 

It is possible to write a similar decomposition also for the Euclidean EPRL vertex amplitude. The derivation is the same as in the Lorentzian case, and it is based on the Cartan decomposition of \spinF{} \eqref{eq:CartanDSpin4}. The EPRL Euclidean vertex amplitude can be expressed as a superposition of \SUT{} $\{15j\}$ symbols weighted by the product of four Euclidean booster functions
\begin{equation}
\label{eq:amplEucl}
A^E_v(j_f,i_e)=\sum_{l_f,k_e}\left(\prod_e d_{k_e}B_4^E(j_f,l_f,i_e,k_e)\right)\{15j\}(l_f,k_e,i')\ .
\end{equation}
The Euclidean booster functions $B_4^E$ are defined in terms of the product of four \spinF{} reduced matrix elements in the $(\gamma j_f, j_f)$ representation contracted with $4jm$-symbols:
\begin{equation}\label{Spin4_Boost}
B^E_4(j_a,l_a,i,k)\equiv\sum_{m_a}
\FJM{j_a}{m_a}^{(i)}
\int_0^{2\pi}d\mu(t)\prod_a d^{(\gamma j_a,j_a)}_{j_al_am_a}(t) 
\FJM{l_a}{m_a}^{(k)}
\ ,
\end{equation}
where $d^{(\gamma j_a,j_a)}$ are the reduced matrix elements \eqref{eq:dSpin4} and $d\mu(t)$ is the part over $E^+$ part of the \spinF{} Haar measure \eqref{eq:HaarSpin4}.


\section{The map between the spin foam models}
\label{sec:map}


\subsection{Mapping algebras and groups}
\label{sec:algebras}


The Lie algebra $\mathfrak{su}(2)$ is a compact real form of \sltc{} \cite{Hall2015}. Therefore, we get the (complex) algebra $\mathfrak{sl}(2, \CC)$ of \sltc{} by complexification
\begin{equation}
  \mathfrak{sl}(2, \CC) = \mathfrak{su}(2) \oplus i \mathfrak{su}(2)\ .
\end{equation}
If we consider the (real) algebra of $Spin(4)$, $\mathfrak{spin}(4) \simeq \mathfrak{su}(2) \oplus \mathfrak{su}(2)$, we get the (realification of the) algebra of \sltc{} by ``rotating'' half of the algebra to purely imaginary generators:
\begin{equation}
  \mathfrak{spin}(4) \simeq \mathfrak{su}(2) \oplus \mathfrak{su}(2) \ \to \ \mathfrak{su}(2) \oplus i \mathfrak{su}(2) \simeq \mathfrak{sl}(2, \CC)\ .
\end{equation}
Equivalently, rotating the generator of boosts in \sltc{} (considering its Lie algebra as a real algebra) we obtain the algebra of \spinF{}. Using the canonical bases for \spinF{} and \sltc{} introduced in the previous sections, the rotation maps the generators of Euclidean boosts to the generators of Lorentzian boosts and vice versa:
%
\begin{equation}
  \label{eq:mapAK}
  (\vec{L}, i\vec{K}) \simeq \mathfrak{spin}(4) \qq{and} (\vec{L}, -i\vec{A}) \simeq \mathfrak{sl}(2, \CC)\ .
\end{equation}
We write these isomorphisms of (real) Lie algebras as $\vec{A} \leftrightarrow i\vec{K}$ and $\vec{K} \leftrightarrow -i\vec{A}$. The map \eqref{eq:mapAK} induces a map between group elements as follows.
The Cartan decomposition $Spin(4) = D \cdot E^+ \cdot D$ we introduced in Section \ref{sec:math-Spin4}, is analogous to the Cartan decomposition $SL(2,\CC) = SU(2) \cdot A^+ \cdot SU(2)$ where by $\cdot$ we mean the group product. Therefore, the map \eqref{eq:mapAK} induces a map between the compact subgroup $E^+$ in \eqref{eq:CartanSpin4} and the non-compact subgroup $A^+$ in \eqref{eq:CartanSL2C} seen as subgroups of the complexified groups $Spin(4)_\CC \simeq SL(2,\CC)_\CC$. For example, the map from $E^+$ to $A^+$ can be achieved by sending 
\begin{equation}\label{eq:mapG}
t\longrightarrow ir
\end{equation}
where $t \in [0\ 2\pi)$ parametrizes $E^+$ and $r \in \RR^+$ parametrizes $A^+$.

This relation between the compact subgroup $E^+$ and the non-compact subgroup $A^+$ can be given an interesting geometrical interpretation. Since the action of $Spin(4)$ is transitive on $SU(2) \simeq S^3 \subset \mathbb{H}$ and $D$ stabilizes the identity, the 3-sphere $S^3$ is a homogeneous space for $Spin(4)$, and we can identify the quotient subgroup $Spin(4)/D$ with the 3-sphere. There is a similar result for \sltc{} (a well known construction of geometric analysis \cite{helgason1984groups}): the quotient group $SL(2,\CC)/SU(2)$ can be identified with hyperbolic 3-space $H^3$.

In light of the Cartan decompositions, write an element of the quotient $Spin(4)/D$ as $de(t)D$, $e(t) \in E^+, d \in D$,  and an element of $SL(2,\CC)/SU(2)$ as $ka(r)K$, $a(r) \in A^+, k \in K = SU(2)$. 
The parameters $r,t$ act as radial coordinates in the corresponding 3-manifolds. Hence the inverse map $r \to -it$ from $A^+$ to $E^+$ can be interpreted geometrically as mapping hyperbolic 3-space to spherical 3-space, similarly to the usual rotation $t \to i\tau$ from physical time to Euclidean time that transforms Lorentzian metrics into Euclidean ones (and in particular flat Minkowski space to flat Euclidean space). This becomes manifest if we consider the metric of hyperbolic 3-space in radial coordinates
\begin{equation}\label{eq:metric-H}
  \dd H^2 = \dd r^2 + \sinh ^2 r\ \dd \Omega_2^2
\end{equation}
where $\dd \Omega_2$ is the metric on the 2-sphere. The map $r \to -it$ maps this metric to 
\begin{equation}\label{eq:metric-S}
  \dd S^2 = \dd t^2 + \sin ^2 t\ \dd \Omega_2^2
\end{equation}
which is exactly the metric of the 3-sphere (up to an innocuous global minus sign).

From the metrics \eqref{eq:metric-H} and \eqref{eq:metric-S} we can also read the Jacobians that enter the Haar measures of \spinF{} and \sltc{} given in Sections \ref{sec:math-SL2C} and \ref{sec:math-Spin4}. The measure on $E^+$ gets mapped to $-$ the measure on $A^+$
\begin{equation}
  \dd\mu(t)=  \frac{1}{\pi} \sin^2t \longrightarrow \  - \frac{1}{\pi} \sinh^2r \ \dd r = -\dd\mu(r) \ .
\end{equation}
and the Haar measure of \spinF \eqref{eq:HaarSpin4} gets mapped to the Haar measure of \sltc{} \eqref{eq:HaarSL2C}
\begin{equation}
  \dd\mu_{Spin(4)} =  \frac{1}{\pi} \sin^2t \ \dd t \ \dd u\  \dd v \  \longrightarrow \  \frac{1}{\pi} \sin^2ir \ \dd r \ \dd u\  \dd v = - \dd\mu_{SL(2,\CC)} \ .
\end{equation}.


\subsection{Mapping representations and matrix elements}
\label{sec:map-matrixel}

The isomorphisms of real Lie algebras \eqref{eq:mapAK} can be used to find a correspondence between the unitary irreducible representations of \spinF{} and \sltc{}. We need the following facts:
\begin{itemize}
\item[(i)] the complexification of $\mathfrak{spin}(4)$ is isomorphic to the complexification of $\mathfrak{sl}(2,\CC)$ (as a real algebra)
\begin{equation}
  \mathfrak{spin}(4)_\CC \,\simeq\, \mathfrak{su}(2)_\CC \oplus \mathfrak{su}(2)_\CC \,\simeq\, \mathfrak{sl}(2,\CC) \oplus \mathfrak{sl}(2,\CC) \,\simeq\, \mathfrak{sl}(2,\CC)_\CC \ ,
\end{equation}
\item[(ii)] for any Lie algebra $\mathfrak{g}$, (real linear) representations of $\mathfrak{g}$ on a complex vector space extend uniquely to holomorphic (i.e. complex linear) representations of $\mathfrak{g}_\CC$ on the same vector space.
\end{itemize}
These two results imply that we can map the $(p,k)$ and $(\rho,k)$ representations working with the complexified algebras. Concretely, we can compute the action of the Casimirs in the complexified algebras and find the map between representations looking at their eigenvalues on the respective canonical bases. From $\vec{A} \leftrightarrow i\vec{K}$ we get
\begin{align}
\vec{A}^2+\vec{L}^2 &\leftrightarrow - (\vec{K}^2-\vec{L}^2) & p^2+k^2 -1&\leftrightarrow - \rho^2 +k^2 - 1 \ ,\\
\vec{A}\cdot\vec{L} &\leftrightarrow  i\vec{K}\cdot \vec{L}& p k &\leftrightarrow  i \rho k  \ .
\end{align}
Looking at the second Casimir we read the map from \sltc{} to \spinF{} representations:
\begin{equation}\label{eq:MAP2}
(\rho \to -ip, k) \simeq (p,k)\ .
\end{equation}
The converse isomorphism $\vec{K} \leftrightarrow -i\vec{A}$ provides the map from \spinF{} to \sltc{}:
\begin{equation}\label{eq:MAP}
  (p \to i\rho, k) \simeq (\rho,k)\ .
  \end{equation}
In the following we also write $(p,k) \leftrightarrow (i\rho,k)$ to denote both \eqref{eq:MAP2} and \eqref{eq:MAP}. These correspondences are defined up to a global minus sign, which is irrelevant since the irreps $(\rho, k)$ and $(-\rho, -k)$ are unitarily equivalent. These maps between representations can be realized explicitly in terms of matrix elements, as follows.
Using analytic continuation of the representation labels, the \sltc{} matrix elements in the $(\rho,k)$ representation can be obtained from the \spinF{} matrix elements in the $(p,k)$ representation using \eqref{eq:MAP} and \eqref{eq:mapG}. The converse from \spinF{} to \sltc{} is also possible. 
This result can be found in the group theory literature \cite{Anderson:1970gq, Basu:1977ii,Rashid:1979xv,Wong:1977}. Each paper in this list use a different technique and a set of different conventions, making challenging to compare them. The proof we present in Appendix \ref{app:proof} is original and it is based only on elementary properties of the hypergeometric functions. Deriving a new proof helps us to be immune to the numerous different conventions that plague the literature on the subject. 

We summarize the result here and refer to Appendix \ref{app:proof} and \cite{TesiAlessandro} for more details for the interested reader. Because of the Cartan decompositions \eqref{eq:CartanDSL2C} and \eqref{eq:CartanDSpin4} it is sufficient to show that the reduced matrix elements of \sltc{} given by \eqref{eq:dSL2C} can be analytically continued in the \spinF{} ones \eqref{eq:dSpin4}. We find that
\begin{equation}\label{eq:dSL2C_CG}
\begin{split}
d_{jlm}^{(\rho,k)}(r)&=\sum_{n}e^{-(i\rho-k-1+m-2n)r}\\
&\braket{\frac{i\rho+k-1}{2},\frac{i\rho-k-1}{2}+m-n\,;\,\frac{i\rho-k-1}{2},n-\frac{i\rho-k-1}{2}}{j,m}\\
&\braket{\frac{i\rho+k-1}{2},\frac{i\rho-k-1}{2}+m-n\,;\,\frac{i\rho-k-1}{2},n-\frac{i\rho-k-1}{2}}{l,m}\\
&+\sum_{n}e^{-(-i\rho+k-1+m-2n)r}\\
&\braket{\frac{-i\rho-k-1}{2},\frac{-i\rho+k-1}{2}+m-n\,;\,\frac{-i\rho+k-1}{2},n-\frac{-i\rho+k-1}{2}}{j,m}\\
&\braket{\frac{-i\rho-k-1}{2},\frac{-i\rho+k-1}{2}+m-n\,;\,\frac{-i\rho+k-1}{2},n-\frac{-i\rho+k-1}{2}}{l,m}\\ 
&
\end{split}
\end{equation}
where $\braket{j_1,m_1,j_2,m_2}{j,m}$ are the analytic continuation to complex spins of \SUT{} Clebsch-Gordan coefficients \eqref{eq:ComplexCG}. Notice that, under the change of sign of the representation labels $(\rho,k) \to (-\rho,-k)$ the first term becomes the second and vice-versa. This is an explicit sign of the unitary equivalence of the representations $(\rho,k)$ and $(-\rho,-k)$.

If we perform the analytic continuation in both the representation label $\rho \to -ip$, equivalently $i\rho \to p$, and the group element $r \to -it$, equivalently $ir \to t$, we obtain  
\begin{equation}
\begin{split}
d_{jlm}^{(-ip,k)}(it)&=\sum_{n}e^{(p-k-1+m-2n)it}\\
&\left\langle\left.\left(\frac{p+k-1}{2},\frac{p-k-1}{2}+m-n\right),\left(\frac{p-k-1}{2},n-\frac{p-k-1}{2}\right)\right|j,m\right\rangle\\
&\left\langle\left.\left(\frac{p+k-1}{2},\frac{p-k-1}{2}+m-n\right),\left(\frac{p-k-1}{2},n-\frac{p-k-1}{2}\right)\right|l,m\right\rangle\\
&+\sum_{n}e^{(-p+k-1+m-2n)it}\\
&\left\langle\left.\left(\frac{-p-k-1}{2},\frac{-p+k-1}{2}+m-n\right),\left(\frac{-p+k-1}{2},n-\frac{-p+k-1}{2}\right)\right|j,m\right\rangle\\
&\left\langle\left.\left(\frac{-p-k-1}{2},\frac{-p+k-1}{2}+m-n\right),\left(\frac{-p+k-1}{2},n-\frac{-p+k-1}{2}\right)\right|l,m\right\rangle \ .\\ 
&
\end{split}
\end{equation}
The second term vanishes identically since $k\leq j,l \leq p-1$ while the Clebsch-Gordan coefficients vanishes if $j,l\leq -p-1$. If we shift the first summation $n\to n'+\frac{p-k-1}{2}$ we obtain the expression for the reduced matrix elements of \spinF{} as in \eqref{eq:dSpin4}. 

For later convenience, we introduce the generalized matrix element function on the complex plane 
\begin{equation}\label{eq:SL2C_SmallD_Expl}
\begin{split}
d^{(a,k)}_{jlm}(z) =&(-1)^{j-l}\sqrt{\frac{\left(a-j-1\right)!\left(j+a\right)!}{\left(a-l-1\right)!\left(l+a\right)!}}\frac{\sqrt{(2j+1)(2l+1)}}{(j+l+1)!}z^{-(a-k-m-1)}\\
&\sqrt{(j+k)!(j-k)!(j+m)!(j-m)!(l+k)!(l-k)!(l+m)!(l-m)!}\\
&\sum_{s,t}(-1)^{s+t}z^{2t}\frac{(k+s+m+t)!(j+l-k-m-s-t)!}{t!s!(j-k-s)!(j-m-s)!(k+m+s)!(l-k-t)!(l-m-t)!(k+m+t)!}\\
&\ _{2}F_{1}\left[\{l-a+1,k+m+s+t+1\},\{j+l+2\};1-z^{2}\right] \ ,
\end{split}
\end{equation}
where $z\in \CC$, $k\in \mathbb{Z}/2$ and $a$ can be either a half integer greater than $1$ or a purely imaginary number. 
We use for \eqref{eq:SL2C_SmallD_Expl} a notation similar to the reduced matrix elements of \sltc{} and \spinF{} and we distinguish it from them by the use of the complex argument.
The slight abuse of notation is justified since, if $a=i\rho$ and $z=e^{-r}$ then \eqref{eq:SL2C_SmallD_Expl} turns into the \sltc{} reduced matrix elements $d^{(i\rho,k)}_{jlm}(e^{-r}) \equiv d_{jlm}^{(\rho,k)}(r)$\footnote{This must not be confused with the analytic continuation $\rho \to -ip$ which could be equivalently written as $i\rho \to p$.}. If $a=p \geq 1$, $p\in \mathbb{Z}/2$ and $z = e^{it}$ then \eqref{eq:SL2C_SmallD_Expl} turns into the \spinF{} reduced matrix elements $d^{(p,k)}_{jlm}(e^{it}) \equiv d_{jlm}^{(p,k)}(t)$.


\subsection{Mapping the vertex amplitude}
\label{sec:map-vertex}
We compare the expressions of the Lorentzian \eqref{eq:amplLor} and the Euclidean \eqref{eq:amplEucl} vertex amplitude. To find a relation between the amplitudes, it is sufficient to study the relation between the Lorentzian booster function \eqref{SL2C_Boost} and the Euclidean booster function \eqref{Spin4_Boost}.
\subsubsection*{Booster functions as integrals}
The generalized matrix element $d^{(a,k)}_{jlm}(z)$ \eqref{eq:SL2C_SmallD_Expl} and its connection with the reduced matrix elements of \spinF{} and \sltc{} induces an embedding of $E^+$ in the complex $z$-plane to the unit circle $e^{it} \in S^1$ and an embedding of $A^+$ to the unit interval $e^{-r}\in[0\ 1]$.

The booster functions \eqref{SL2C_Boost} and \eqref{Spin4_Boost} depend on the integration of the product of four reduced matrix elements over the appropriate subgroup. This translates into the integration of the generalized matrix element $d^{(a,k)}_{jlm}(z)$ \eqref{eq:SL2C_SmallD_Expl} in the complex plane along the unit circle in the Euclidean case, or the unit interval in the Lorentzian case. In the following, we relate the integrals of products of $d^{(i\rho, k)}(z)$ along these two paths.

For simplicity, we will focus on the minimal case $j = l = k$. This simplification allows us to avoid the clutter of the additional sums over the indices $s,t$ present in the reduce matrix elements in the non-minimal case. However, the calculation in the general case follows the same steps and we can apply the same arguments of the simplified case. It is just more cumbersome and confusing to keep track of all the terms. In the simplified case the matrix elements become 
\begin{equation}
  d^{(a,k)}_{kkm}(z) = z^{(k+m+1)-a}\ _{2}F_{1}\left[\{k-a+1,k+m+1\},\{2k+2\};1-z^{2}\right] \ .
\end{equation}
The integrand of \eqref{SL2C_Boost} and \eqref{Spin4_Boost} with general $a$ and omitting the intertwiner index for brevity is
\begin{equation}
\label{eq:f}
\begin{split}
f(a_i,k_i; z) &= \sum_{m_i}\FJM{k_i}{m_i} \prod_{i=1}^{4}  d^{(a_i,k_i)}_{k_{i} k_{i} m_{i}}(z) \FJM{k_i}{m_i}\\
&=\sum_{m_i}  z^{K-A+4+M}  \FJM{k_i}{m_i} \prod_{i=1}^{4} \  _{2}F_{1}\left[\{k_i-a_i+1,k_i+m_i+1\},\{2k_i+2\};1-z^{2}\right] \FJM{k_i}{m_i}
\end{split}
\end{equation}
with $K = \sum_i k_i$, $A = \sum_i a_i$, $M=\sum_i m_i$. Notice that the $4jm$-symbols vanish if $M\neq 0$, therefore in the following we will assume $M = 0$ if needed. 

\subsubsection*{From Euclidean to Lorentzian integrals}

The Euclidean booster functions are defined as the integral of \eqref{eq:f} with $a_i = p_i = \gamma j_i$ half integer greater than $1$. Our results are valid independently from the imposition of the $Y_\gamma$ map that constrains $p_i$ to assume specific values.  In this section, we will keep $p$ (and $\rho$ in the Lorentzian case) generic until the very end. To avoid unnecessary confusion, we will call the integral of \eqref{eq:f} $I_E(p_i, k_i)$, emphasizing its dependence on \spinF{} labels $p_i$ and $k_i$, and discuss the connection to the booster function later. The \spinF{} Haar measure induces on the complex plane the integration measure  
\begin{equation}
 \dd t\ \frac{\sin^2(t)}{\pi}\quad \longrightarrow \quad  i\ \frac{(1-z^2)^2}{4\pi z^2} \frac{\dd z}{z} = \dd\mu(z) \ .
\end{equation}
The Euclidean integral $I_E$ expressed as a contour integral over the unit circle is 
\begin{equation}
  \begin{split}
  I_E(p_i, k_i) = \frac{i}{4\pi} \oint_{S^1}\dd z\  (1-z^2)^2 &\sum_{m_i}\FJM{k_i}{m_i} z^{K-A+M+1} \ \times \\
   &\prod_{i=1}^4 \  _{2}F_{1}\left[\{k_i-p_i+1,k_i+m_i+1\},\{2k_i+2\};1-z^{2}\right] \FJM{k_i}{m_i}\ .
  \end{split}
\end{equation}
The first argument of all the hypergeometric functions $k_i - p_i$ is a strictly negative integer since $k_i < p_i$ (see Section~\ref{sec:math-Spin4}). It follows that the hypergeometric functions reduce to polynomials in $1-z^2$. However, from $k_i - p_i + 1 = -2J_{Ri}$ we find that the prefactor $z^{K-A+M+1}$ introduces a pole singularity in $z = 0$ and the complete integrand is meromorphic. The integral can be evaluated using Cauchy residue theorem, and the result is $2\pi i$ times the residue at $z=0$ which depends non-trivially on all the parameters through the product of the hypergeometric functions.

\medskip

As an example, we work out explicitly the degenerate case $j_{Ri} = 0$ where it is particularly simple to compute the integral $I_E$ with Cauchy's closed curve theorem. Here $k_i=p_i-1$ and all the hypergeometric functions are equal to $1$. We can take $M=0$ otherwise the $4jm$-symbols vanish, so that $K-A=-4$ in this case. The function $(1-z^2)^2 z^{-3}$ has a pole with residue $-2$ in the origin. The integral is immediately calculable, reintroducing the intertwiner indices $(a)$ and $(b)$ explicitly
\begin{equation}
  \label{eq:eucl-integral-simple}
\sum_{m_i}\FJM{k_i}{m_i}^{(a)}  \frac{i}{4\pi} 2\pi i (-2) \FJM{k_i}{m_i}^{(b)} =  \frac{\delta_{ab}}{2a+1} \ .
\end{equation}
This result is exactly what one would expect by performing the integral in the canonical basis remembering that $D^{J_{Ri}}(g_R)=1$ if $J_{Ri}=0$ for any $g_R\in SU(2)$.

\medskip

\begin{figure}[h]
  \centering
 \includegraphics{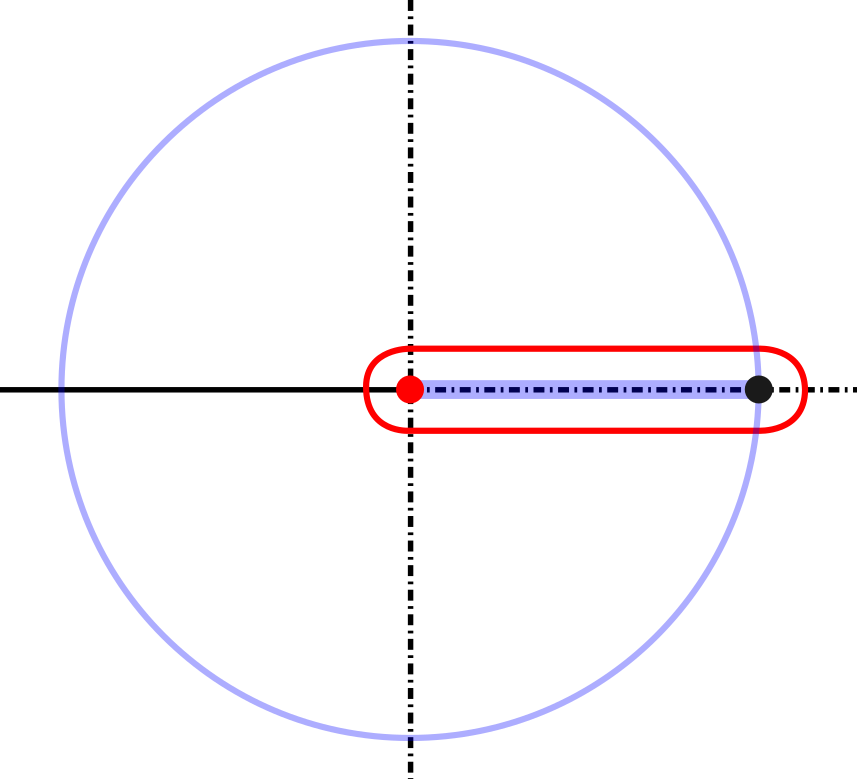} 
 \caption{The contour $\conte$ in red. The point $z=0$ in red is a pole singularity in the Euclidean case and a branch point singularity in the Lorentzian case. The black point is $z=1$. In the Lorentzian case the integrand is discontinuous on the punctured axes.}
 \label{fig:contour}
\end{figure}

The integrand is analytic in the punctured plane. Therefore, it is actually irrelevant which contour one uses, as long as it contains $z=0$. Let us consider the contour $\conte$ represented in red in Figure \ref{fig:contour}. The horizontal segments have small distance $\varepsilon$ from the real axis. The semicircles around $0$ and $1$ have small radius $\varepsilon$. 

We can deform the contour integral in \eqref{eq:BoosterSpin4z} from the unit circle to the contour $\conte$ obtaining a completely equivalent definition of the Euclidean integral. Furthermore, taking the limit $\varepsilon \to 0$ does not change the result. We will base the analytic continuation of the Euclidean integral on this expression:
\begin{equation}
\label{eq:BoosterSpin4z}
\begin{split}
  I_E(p_i, k_i) = \lim_{\epsilon\to 0}\frac{i}{4\pi} \oint_{\conte}\dd z\  (1-z^2)^2 &\sum_{m_i}\FJM{j_i}{m_i} z^{K-A+M+1}  \ \times \\
    &\prod_{i=1}^4 \  _{2}F_{1}\left[\{k_i-a_i+1,k_i+m_i+1\},\{2k_i+2\};1-z^{2}\right] \FJM{j_i}{m_i} \ .
\end{split}
\end{equation}
The Euclidean booster function is obtained from \eqref{eq:BoosterSpin4z} by imposing a restriction on $p=\gamma k$ coming from the $Y_\gamma$-map. 

\medskip

The Lorentzian booster functions are defined as the integral of \eqref{eq:f} with $a_i =  i \rho_i$ with $\rho_i$ a positive real number. Once again we postpone the imposition of the $Y_\gamma$-map to the very end and to avoid unnecessary confusion we will call the integral $I_L(\rho_i, k_i)$, emphasizing its dependence on \sltc{} labels $\rho_i$ and $k_i$. The \sltc{} Haar measure induces on the complex plane the integration measure
\begin{equation}
 \dd r\ \frac{\sinh^2(r)}{\pi} \quad \longrightarrow \quad  - \frac{(1-z^2)^2}{4 \pi z^2} \frac{\dd z}{z} = i \dd\mu(z)\ .
\end{equation}
The Lorentzian integral $I_L$ can be expressed as a line integral over the unit interval as 
\begin{equation}
\label{eq:BoosterSl2Cz}
\begin{split}
  I_L(\rho_i, k_i) = \frac{1}{4\pi} \int_{[0\ 1]}\dd z\ &\sum_{m_i}\FJM{j_i}{m_i} (1-z^2)^2 z^{K-A+M+1}\ \times \\
   &\prod_{i=1}^4 \  _{2}F_{1}\left[\{k_i-i\rho_i+1,k_i+m_i+1\},\{2k_i+2\};1-z^{2}\right] \FJM{j_i}{m_i}\ .
\end{split}
\end{equation}
The integrand of \eqref{eq:BoosterSl2Cz} differs from the Euclidean one \eqref{eq:BoosterSpin4z} only by a factor of $i$ and the different values of $a_i$ and $A=\sum_i a_i = i \sum_i \rho_i$. This however implies that the Lorentzian integrand is not meromorphic anymore. The hypergeometric functions develop a branch point singularity in $1-z^2=1$, i.e. $z=0$. The analytic continuation of the hypergeometric series outside of the unit disc using Euler's formula has a branch cut discontinuity along real numbers $x \geq 1$ \cite{NIST:DLMF}. In our case, the hypergeometric function is computed in $1-z^2$. The branch cut discontinuity is along the whole imaginary axis, represented as a punctured line in Figure \ref{fig:contour}, and there are two disconnected domains of analyticity. We assign the principal branch $|\arg(z)| < \pi/2$ on both sides of the imaginary axis and we define the value on the imaginary axis minus the origin by continuity from the left. 

At the origin each one of the hypergeometric functions in \eqref{eq:f} is in general divergent. This happens when $\Re(\delta_i-\alpha_i-\beta_i)<0$ where $\alpha_i$, $\beta_i$, $\delta_i$  are the three parameters of $\  _{2}F_{1}\left[\{\alpha_i,\beta_i\},\{\delta_i\};1-z^{2} \right]$. In our case $\Re(\delta_i-\alpha_i-\beta_i) = -m_i$, therefore for some $m_i$ the hypergeometric function is divergent at most of order $d_i = 2\max(0, m_i)$, i.e. 
\begin{equation}
\label{eq:z1property}
  \lim_{z \to 0^-} z^{d_i+\sigma} {}_2F_1\left[\{k_i-a_i+1,k_i+m_i+1\},\{2k_i+2\};1-z^{2}\right] = 0
\end{equation}
for any real $\sigma > 0$
\footnote{The hypergeometric function has the property \cite{NIST:DLMF}
\begin{equation}
\lim_{w\to 1_{-}}(1-w)^{\alpha+\beta-\delta} {}_2F_1[\{\alpha,\beta\},\{\delta\},w] = \frac{\Gamma(\delta) \Gamma(\alpha+\beta-\delta)}{\Gamma(\alpha) \Gamma(\beta)}\ ,
\end{equation}
therefore 
\begin{equation}
\lim_{w\to 1_{-}}(1-w)^{\alpha+\beta-\delta+\sigma} {}_2F_1[\{\alpha,\beta\},\{\delta\},w] = 0\ ,
\end{equation}
for any real $\sigma >0$. Substituting $w \to 1-z^2$ we obtain \eqref{eq:z1property}.}. The product of four hypergeometric functions in \eqref{eq:f} is divergent at most logarithmically in the origin, i.e. $\sum_i d_i = 2\max(0, \sum_i m_i)= 2\max(0, M)=0$, since we can always take $M=0$. This implies that the prefactor $z^{K-A+1}$ cures any potential divergence in the origin of \eqref{eq:f}.\footnote{
Notice that this is not peculiar for the minimal case $j_i=l_i=k_i$. In the general case the order of divergence of the product of the four hypergeometric functions is given by $2\max(0,m_i + t_i-(j_i- k_i  - s_i))$. The summation over $s_i$ is such that $j_i- k_i  - s_i$ is always positive, the $z^{2t_i}$ factor counterbalances the possible divergence of order $2t_i$ and $\sum_i m_i=0$. We conclude that also in the general case the divergence (at most logarithmic) of the product of hypergeometric functions is cured by the prefactor $z^{K-A+1}$.}

The branch cut discontinuity along the imaginary axis is not the only one. The prefactor $z^{K-A+1}$ also has branch point singularities in 0 and $\infty$ since $A$ is purely imaginary. For this term we consider the branch $|\arg(-z)|<\pi$ so that the discontinuity is along the positive real axis (represented in Figure~\ref{fig:contour} as a punctured line). We note, however, that the hypergeometric functions are continuous across the positive real axis.

\medskip 

We conclude our analysis by relating the integral on the unit interval in \eqref{eq:BoosterSl2Cz} to the contour integral over $\conte$ of the same function in the limit of $\varepsilon \to 0$. The latter is $i$ times the analytic continuation of the Euclidean integral \eqref{eq:BoosterSpin4z} to purely imaginary $a_i$.

We split the contour $\conte$ in four pieces: let $\conte^0$ be the small semicircle around 0, $\conte^1$ be the small semicircle around 1, $\conte^+$ be the straight line above the real axis and $\conte^-$ be the straight line below the real axis. The integral along $\conte^1$ vanishes since $\Re(K-A+1) > 0$ and the hypergeometric function is regular there. The integral along $\conte^0$ vanishes for the same reason, since the eventual logarithmic divergence at $z = 0$ of the product of the four hypergeometric functions is more than canceled by the prefactor $z^{K-A+1}$. The integrals along $\conte^+$, $\conte^-$ differ by a factor $\exp 2\pi i (K-A+1)$ due to the presence of the branch cut of $z^{K-A+1}$ while the hypergeometric function is continuous in the right half-plane $\Re z > 0$. In the limit $\varepsilon\to 0$ we have
\begin{equation}
  \begin{split}
  \lim_{\varepsilon\to 0}\oint_{\conte} \dd z\ f(a_i; z) &= \lim_{\varepsilon\to 0}\left( \oint_{\conte^+}\dd z\ f(a_i; z) + \oint_{\conte^-}\dd z\ f(a_i; z) \right) \\
  &= \left(e^{2\pi i (K-A+1)} - 1\right) \int_{[0\ 1]}\dd z\  f(a_i; z)
  \end{split}
\end{equation}
taking into account the opposite directions induced by the contour $\conte$. From the definition \eqref{eq:BoosterSpin4z} of the Euclidean integral we conclude that taking its analytic continuation $p_i \to i\rho_i$ we obtain (up to a multiplicative factor) the Lorentzian integral
\begin{equation}
  \label{eq:cont-1}
I_E(p_i, k_i) \ \xrightarrow{\ p_i \to i\rho_i\ }\  i \left(e^{2\pi i \sum_i (k_i - i\rho_i)} - 1\right) I_L(\rho_i, k_i)
\end{equation}
or equivalently
\begin{equation}
  \label{eq:cont-iE-to-iL}
  I_L(\rho_i, k_i) = \frac{i}{1-e^{2\pi i \sum_i (k_i - i\rho_i)}}\ I_E(i\rho_i, k_i)
\end{equation}
where on the right side we mean the analytic continuation of $I_E(p_i, k_i)$ as a function of $p_i$ and $k_i$ to purely imaginary values of the first argument\footnote{More rigorously, the ``analytic continuation'' of a function defined on the integers cannot be performed in the mathematical sense since $\mathbb{Z}$ is not an open subset of $\CC$. Hence one can just define $I_E(i\rho_i, k_i)$ to be the evaluation of $I_E(p_i, k_i)$ with $p_i$ purely imaginary. Notice, however, that in light of the converse result \eqref{eq:cont-iL-to-iE} we can speak unambiguously of the unique function on the whole complex plane that extends both $I_E(p_i, k_i)$ and $I_L(\rho_i, k_i)$ at the same time.}.

\medskip

In deriving \eqref{eq:cont-1} we considered the minimal case $l_i = j_i=k_i$ for simplicity. The generalization to any $j_,l_i\geq k_i$, and in particular to the half-minimal case relevant for the booster function $k_i=j_i$ and $l_i\geq j_i$, is straightforward. All the arguments we made are immediately extended. In particular, the prefactor $z^{K-A+1}$ remains the same and all the considerations about the product of the hypergeometric functions with minimal arguments apply also to the more complicated sum over $s_i,t_i$ of products of hypergeometric functions.

\subsubsection*{From Lorentzian to Euclidean integrals}

Formulae \eqref{eq:cont-1} and \eqref{eq:cont-iE-to-iL} provide also the converse result for the rotation $i\rho_i \to p_i$:
\begin{equation}
  \label{eq:cont-iL-to-iE}
  I_E(p_i, k_i) = i \lim_{q_i \to p_i} \left(e^{2\pi i \sum_i (k_i - q_i)}-1 \right) I_L(-i q_i, k_i) \ ,
\end{equation}
where first we do the analytic continuation $\rho_i \to -iq_i$ of $I_L(\rho_i, k_i)$ with $q_i \in \RR \setminus \mathbb{Z}$ and then we take the limit $q_i \to p_i = k_i + n_i$ and $n_i \in \mathbb{N}^+$ to regularize the product of the vanishing prefactor with the divergent function $I_L(-i q_i, k_i)$. In fact, the defining integral representation \eqref{eq:BoosterSl2Cz} of $I_L(\rho_i, k_i)$ is divergent if we perform the substitution $i\rho_i \to p_i$, for any half-integer $p_i > k_i \geq 0$. However, we can overcome this difficulty noticing that the same apparent obstruction appears for example in the Euler's integral representation of the standard Gamma and Beta functions \cite{NIST:DLMF}. In particular, after the substitution $i \rho_i \to p_i$ the integral in \eqref{eq:BoosterSl2Cz} reduces to
\begin{equation}
  \int_0^1 \dd z\ (1-z^2)^2 z^{1+\sum_i (k_i-p_i)} P[1-z^2]
\end{equation}
where $P[1-z^2]$ stands for a generic polynomial in the variable $1-z^2$ and where we set $M=0$. Changing variables $z^2 \to w$ we can write this integral as a finite sum of Beta functions 
\begin{equation}
  \label{eq:beta-sum}
  \int_0^1 \dd w\ w^{-2 + \frac{1}{2}\sum_i (k_i-p_i+1)} P'[1-w] \sim \sum_j B \left( -2 + \frac{1}{2}\sum_i (k_i-p_i+1), n_j \right)
\end{equation}
with first argument always a negative integer and second argument a positive integer. The Beta function can be analytically continued to complex values of its arguments using for example the Pochammer contour, possibly with simple poles at the negative integers. Since $I_L(-iq_i, k_i)$ tends to \eqref{eq:beta-sum} continuously for $q_i \to p_i$, this implies that it is possible to analytically extend $I_L(\rho_i, k_i)$ to generic complex values of the first parameter, again possibly with simple poles at the negative integers, i.e. at the values $k_i-p_i+1$ relevant for our case. The simple form \eqref{eq:beta-sum} holds however only in small a neighborhood of the poles since in general the hypergeometric functions won't be expressible as simple polynomials. Remarkably, the vanishing prefactor in \eqref{eq:cont-iL-to-iE} exactly cancels the divergence of the analytically continued $I_L(-iq_i, k_i)$ at its simple poles. 

We show how this works in the simple case $p_i = k_i + \sigma, \sigma \to 1$ where we expect to recover \eqref{eq:eucl-integral-simple}. In this case the hypergeometric functions are equal to unity and the integral in \eqref{eq:BoosterSl2Cz} reduces to
\begin{equation}
  \int_0^1 \dd z\ (1-z^2)^2 z^{1-4\sigma}.
\end{equation}
The integral is divergent for $\sigma \to 1$. Changing variables $z^2 \to w$ we obtain the integral
\begin{equation}
  \frac{1}{2} \int_0^1 \dd w\ (1-w)^2 w^{-2\sigma} = \frac{1}{2} B(1-2\sigma, 3)
\end{equation}
which as a complex function of $\sigma$ has a simple pole in $\sigma = 1$ with residue $-1$. Recalling the prefactor from \eqref{eq:cont-iL-to-iE} we get
\begin{equation}
  \frac{i}{2} \lim_{\sigma \to 1} \left( e^{- 8\pi i \sigma} -1 \right) B(1-2\sigma, 3) = 4 \pi
\end{equation}
and inserting this in \eqref{eq:BoosterSl2Cz} we get exactly the Euclidean result \eqref{eq:eucl-integral-simple}.

\medskip 


Notice that since each $\rho_i$ span an open subset of $\CC$, the previous considerations support strongly the conjecture that the two functions $I_E(p_i, k_i)$ and $I_L(\rho_i, k_i)$ are particular integral representations of a unique function $I(a_i, k_i)$ defined on the whole space $\CC^4 \times \mathbb{Z}^4$, which agrees with $I_E(p_i, k_i)$ for $a_i = p_i$ and with $I_L(\rho_i, k_i)$ for $a_i=i \rho_i$. Hence, we can speak unambiguously of \emph{the} analytic continuation of $I_E$ and $I_L$. We do not provide a rigorous proof of this interesting claim here, which would require a more careful treatment of the interplay between the analytic continuation of the hypergeometric functions, the prefactor $z^{K-A+1}$ and the integration on the unit interval.

\subsubsection*{Analytic continuation of vertex amplitudes}

Imposing the $Y_\gamma$ map to both the Lorentzian and Euclidean integrals we find the desired relation between the (analytic continuations of the) Lorentzian and Euclidean booster functions. The analytic continuation $p_i \to  i \rho_i$ reads $\gamma j_i \to i \gamma j_i$ and can be interpreted fascinatingly as the rotation of the Immirzi parameter $\gamma\to i \gamma$
\begin{equation}
  B^L_4(j_a,l_a,i,k) = \frac{i}{1-e^{2\pi(i+\gamma)\sum_i j_i }} B_4^E(j_a,l_a,i,k)^{(\gamma \to i \gamma)} \ .
\end{equation}
The prefactor can be furthermore simplified remembering that $\sum_i j_i$ is always an integer, obtaining the simpler
\begin{equation}
\label{eq:LorFromEucl}
  B^L_4(j_a,l_a,i,k) = \frac{i}{1-e^{2\pi\gamma\sum_i j_i }} B_4^E(j_a,l_a,i,k)^{(\gamma \to i \gamma)} \ .
\end{equation}
From the definition of the vertex amplitudes \eqref{eq:amplLor} and \eqref{eq:amplEucl} and using \eqref{eq:LorFromEucl} we find
\begin{equation}
A_v^L(j_f,i_e) = \left( \prod_e \frac{i}{1-e^{2\pi\gamma\sum_{i\in e} j_i }} \right) A_v^E (j_f,i_e)^{(\gamma \to i \gamma)}  \ .
\end{equation}
Vice versa, the analytic continuation $i\rho_i \to p_i$ when the $Y_\gamma$ map is imposed reads $i \gamma j_i \to \gamma j_i$ and can be interpreted as the inverse rotation of the Immirzi parameter $i\gamma \to \gamma$, equivalently $\gamma \to -i\gamma$. The booster functions are related by
\begin{equation}
      B_4^E(j_a,l_a,i,k)  = i \left[ \left(e^{2\pi\gamma\sum_i j_i }-1 \right) B^L_4(j_a,l_a,i,k) \right]^{(i \gamma \to \gamma)}
\end{equation}
where the rotation $i\gamma \to \gamma$ must be regularized taking the limit defined in \eqref{eq:cont-iL-to-iE}.
A similar expression for the vertex amplitude is readily obtained
\begin{equation}
A_v^E (j_f,i_e) = i \left[ \prod_e \left( e^{2\pi\gamma\sum_{i\in e} j_i }-1 \right) A_v^L(j_f,i_e) \right]^{(i \gamma \to \gamma)} \ .
\end{equation}
This completes the derivation of the prescription for the analytic continuation of Euclidean and Lorentzian vertex amplitudes one into the other.


\section{Conclusion}
\label{sec:conclusion}


In the spin foam literature, the Euclidean and Lorentzian EPRL models are traditionally presented differently. Although the guiding principle is the same, that is imposing the linear simplicity constraints weakly, the resulting implementations look dissimilar. 
Performing calculations in the Euclidean model is much more straightforward than in the Lorentzian one. Therefore, many results are derived within the first \cite{Bianchi:2009ri, Barrett:2009gg, Hellmann:2013gva, Bahr:2015gxa} and then inferred to be valid in the second or re-derived from scratch \cite{Bianchi:2011hp, Barrett:2009mw}. Moreover, the model with Euclidean signature carries the stigma of not being relevant or connected to physical calculations.

\medskip

When formulated in the canonical basis and adopting the Cartan decomposition, the two models look very alike, differing only in the booster functions \eqref{eq:BoosterSl2Cz} and \eqref{eq:BoosterSpin4z}, thanks to a few key correspondences. First, the algebra of \sltc{} maps to the algebra of \spinF{} if we rotate the generators of the Lorentzian boosts into $i$ times the generators of the Euclidean ``boosts'' $\vec{A} \leftrightarrow i\vec{K}$ and vice versa. Second, this map induces a correspondence between \sltc{} and \spinF{} group elements that, using the Cartan decompositions \eqref{eq:CartanDSL2C} and \eqref{eq:CartanDSpin4}, reduces to a rotation of the non-compact coordinate $ir \to t$.  
Third, the map between the algebras induces a correspondence between irreducible representation $(p,k)$ of \spinF{} and the unitary irreducible representation $(\rho,k)$ in the principal series of \sltc{} as $(p,k) \leftrightarrow (i\rho,k)$.
We obtain the matrix elements in the $(p,k)$ representation of \spinF{} from the matrix elements in the $(\rho,k)$ representation of \sltc{} through the analytic continuation of the representation labels and group elements simultaneously. Finally, we show that the booster functions of the models with different signature, and, more in general, the vertex amplitudes, can be obtained from one another by rotation of the Immirzi parameter $\gamma \leftrightarrow i\gamma$.

\medskip

This work bridges the gap between the spin foam EPRL models with different signatures and gives a prescription on how to map the results obtained in the Euclidean framework to the Lorentzian one and back. 

\medskip

In addition, we found that the integral forms of the booster functions can be thought as special cases of a general integral defined for complex $a_i$ that analytically continues both the Euclidean and Lorentzian integrals. In terms of the $Y_\gamma$ maps that enforce $a_i = \gamma j_i$, we can think of this as an extension of the booster functions to the case of an arbitrary complex Immirzi parameter\footnote{Amusingly, the key contribution of Giorgio Immirzi (who extended an idea from Fernando Barbero) was to highlight how the complex phase space of Ashtekar variables could be canonically transformed to a \emph{real} phase space using a \emph{real} parameter, namely the Immirzi parameter, to avoid the imposition of the reality conditions. }. 
Then, it could be possible in principle to define a general spin foam model defined with complex Immirzi parameter $\gamma$ that reduces to the Lorentzian EPRL model and the Euclidean EPRL model for purely imaginary or real integer values of $\gamma$. This idea is supported by the fact that the $(\rho,k)$ representations of the principal series of \sltc{} are defined for any $\rho \in \CC$, but they are unitary only when $\rho\in\RR$ \cite{naimark1964linear}. 
However, the physical meaning of these hypothetical ``complex EPRL models'' is not clear to us and we leave the exploration of these ideas to future works. 

\medskip

It is interesting to relate our findings to the early formulation of Loop Quantum Gravity. The original canonical formulation of LQG was based on complex (self-dual) Ashtekar variables. In terms of these variables the constraints of the Hamiltonian formulation of General Relativity are low order polynomials. The major drawback is that one has to impose ``reality conditions'' on the canonical variables to recover real Lorentzian GR. Since the quantization of these reality conditions is problematic, the focus of the LQG community has shifted to the use of real variables (the Barbero-Immirzi variables) as soon as they were introduced. The price to pay is a more complicated form of the constraints and a less clear geometric interpretation of the real connection \cite{Samuel:2000ue}. However, in the Euclidean signature these problems do not arise since the self-dual connection is real, and it is possible to show that a ``Wick rotation'' maps 
the Euclidean constraints of General Relativity to the Lorentzian ones \cite{Ashtekar:1995qw,Thiemann:1995ug,Varadarajan:2018uaj}. In this work we define a similar ``Wick rotation'' in the quantum theory using the covariant formulation of Loop Quantum Gravity.

\medskip

Our work opens the way to many interesting ideas that deserve future explorations.
In the context of computer simulations, we expect our result to contribute to numerical codes for the EPRL model \cite{Dona:2018nev, SLNEXT} to speed up the calculation of the booster functions. Alternatively, we could rethink the entire numerical calculation scheme, avoiding the Cartan decomposition, and setting up the computation using the canonical basis of \spinF{} with the rotation $\gamma \to i\gamma$. It would be interesting also to study a possible connection to other analytic continuations of the EPRL spin foam models based on the complexification of the integration domain \cite{Han:2021rjo} or using Markov Chain Monte-Carlo computations \cite{Han:2020npv}. The computation of the Bekenstein-Hawking entropy of black holes in LQG can be derived using state counting after analytically continuing the formula to $\gamma = \pm i$ \cite{Frodden:2012dq,BenAchour:2016mnn}. We can look for an interpretation of this analytic continuation using our results. The applications go beyond LQG and spin foam models. For example, the booster functions are related to the Clebsch-Gordan coefficients of the respective groups \cite{Speziale:2016axj}. We can use our results to relate the Clebsch-Gordan coefficients of \sltc{} in the principal series to the analytic continuation of the ones of \spinF{} (given by a $\{9j\}$ symbol) \cite{Anderson:1970gq,Wong:1977}. One possibility is to use well known explicit formulae for the \sltc{} Clebsch-Gordan coefficients \cite{Kerimov:1978wf} and the $\{9j\}$ symbol \cite{Yutsis2} in terms of sums of products of hypergeometric functions.

\medskip

We conclude with the remark that our prescription for mapping through analytic continuation the unitary irreducible representations of \spinF{} and \sltc{} can be immediately adapted to other spin foam models based on the same gauge groups. More generally, it would be interesting to study possible physical implications of this intriguing analytic continuation beyond the context of spin foam models.


\section{Acknowledgments}


The work of P.D. is partially supported by the grant 2018-190485 (5881) of the Foundational Questions Institute and the Fetzer Franklin Fund. A.N. acknowledges support from the Universit\`{a} di Bologna and its Erasmus+ program.

\begin{appendices}


\section{$SU(2)$ conventions}
\label{app:SU2}


Here we summarize the $SU(2)$ objects we use in the paper. A useful parametrization of a group element $g\in SU(2)$ is the so called Euler angles parametrization
\begin{equation}
g= e^{-i\phi \frac{\sigma_3}{2}} e^{-i\theta \frac{\sigma_2}{2}} e^{-i\psi \frac{\sigma_3}{2}}  \ ,
\end{equation}
where $\sigma_i$ are the Pauli matrices and $0\leq \phi < 2\pi$, $0\leq \theta < \pi$, and $0\leq \psi < 4\pi$. The Haar measure in this parametrization is given by 
\begin{equation}
\label{eq:SU2haar}
dg = \frac{1}{16\pi^2} \sin\theta d\theta d\phi d\psi \ .
\end{equation}
The matrix elements of a group element $g\in SU(2)$ in the representation of spin $j$ is called \textit{Wigner matrix} and in the basis $\ket{j,m}$ is given by:
\begin{align}
\label{eq:WignerSU2}
D^j_{mn}(g)\equiv \bra{j,m}g\ket{j,n}=& e^{i\phi \frac{m}{2}}e^{-i\psi \frac{n}{2}} \sqrt{(j+m)!(j-m)!(j+n)!(j-n)!} \cdot\\
& \sum _{s} \left(\frac {(-1)^{s}\left(\cos {\frac {\theta }{2}}\right)^{2j+n-m-2s}\left(\sin {\frac {\theta }{2}}\right)^{m-n+2s}}{(j+n-s)!s!(m-n+s)!(j-m-s)!}\right) \ .
\end{align}
Which satisfies the orthogonality relation:
\begin{equation}
\int_{SU(2)}dg\overline{D^j_{mn}(g)}D^{j'}_{m'n'}=\frac{1}{2j+1}\delta_{jj'}\delta_{mm'}\delta_{nn'} \ ,
\end{equation}
and the symmetry property:
\begin{equation}
D^j_{mn}(g^\dagger)=\overline{D^j_{nm}(g)}=(-1)^{n-m}D^j_{-n,-m}(g) \ .
\end{equation}
The tensor product of two SU(2) representations $j_1$ and $j_2$ can be decomposed in terms of a sum of $SU(2)$ representations $j$ with $j=|j_1-j_2|,\cdots,j_1+j_2$. The Clebsh-Gordan coefficients
\begin{equation}
\bra{j_1,m_1,j_2,m_2}\ket{j,m}
\end{equation}
relates the states of the three representations. The Clebsh-Gordan coefficients are real and non-zero if and only if:
\begin{equation}
|j_1-j_2|\leq j\leq j_1+j_2\qquad \text{and} \qquad m=m_1+m_2 \ .
\end{equation}
They satisfy the orthogonality relation
\begin{equation}
\sum_{m_1,m_2} \bra{j_1,m_1,j_2,m_2}\ket{j,m}  \bra{j_1,m_1,j_2,m_2}\ket{l,n} =\delta_{jl}\delta_{mn} \ .
\end{equation} 
There are many explicit expressions for the Clebsh-Gordan coefficients. In Section \ref{sec:map-matrixel} we used the Van Der Waerden's formula \cite{Yutsis:1962vcy}:
\begin{equation}
\label{VanDerWa}
\begin{split}
\bra{j_1,m_1,j_2,m_2}\ket{j,m}&=\delta_{m,m_{1}+m_{2}}\sqrt{2j+1}\sqrt{\frac{(j_1+j_2-j)!(j_1-j_2+j)!(-j_1+j_2+j)!}{(j_1+j_2+j+1)!}}\\
&\sqrt{(j_1+m_1)!(j_1-m_1)!(j_2+m_2)!(j_2-m_2)!(j+m)!(j-m)!}\\
&\sum_{t}(-1)^{t} \frac{1}{t!(j_1+j_2-j-t)!(j_1-m_1-t)!(j_2+m_2-t)!(j-j_2+m_1+t)!(j-j_1-m_2+t)!}
\end{split}
\end{equation}
Where the range of summation is given by the existence conditions of the factorials. This expression can be also used to define an analytic continuation of the Clebsh-Gordan coefficients with complex spins \cite{Rashid:1979xv}. A symmetric equivalent of the Clebsh-Gordan coefficients are the Wigner $3jm$-symbols
\begin{equation}
\left(\begin{matrix}
j_1& j_2& j_3\\
m_1& m_2&m_3
\end{matrix}\right)\equiv\frac{(-1)^{j_1-j_2-m_3}}{\sqrt{2j_3+1}}C^{j_3,-m_3}_{j_1m_1j_2m_2}
\end{equation}
The $3jm$-symbols we use are reals and non zero if and only if 
\begin{equation}
|j_1-j_2|\leq j_3\leq j_1+j_2\qquad \text{and} \qquad m_1+m_2 + m_3 =0 \ .
\end{equation}
They satisfy the orthogonality relations:
\begin{align}
\sum_{j,m}(2j+1)\left(\begin{matrix}
j_1& j_2& j\\
m_1& m_2& m
\end{matrix}\right)
\left(\begin{matrix}
j_1  & j_2  & j\\
n_1 & n_2 & m
\end{matrix}\right)&=\delta_{m_1n_1}\delta_{m_2n_2} \ ,\\
\sum_{m_1,m_2}(2j+1)\left(\begin{matrix}
j_1& j_2& j\\
m_1& m_2& m
\end{matrix}\right)
\left(\begin{matrix}
j_1& j_2& l\\
m_1& m_2 &n
\end{matrix}\right)&=\delta_{jl}\delta_{mn} \ .
\end{align}
The integral of thee matrix elements is given by the product of two $3jm$-symbols
\begin{equation}
\int_{SU(2)}dg D^{j_1}_{m_1n_1}(g)D^{j_2}_{m_2n_2}(g)D^{j_3}_{m_3n_3}(g)=
\left(\begin{matrix}
j_1& j_2& j_3\\
m_1& m_2&m_3
\end{matrix}\right)
\left(\begin{matrix}
j_1& j_2& j_3\\
n_1& n_2&n_3
\end{matrix}\right) \ .
\end{equation}

We can couple four $SU(2)$ representations $j_1$, $j_2$, $j_3$, and $j_4$ in many (equivalent) ways. If choosing the recoupling basis $(12)$ we define the $4jm$-symbols as
\begin{equation}\label{4jm}
\left(\begin{matrix}
j_1 & j_2 & j_3 & j_4\\
m_1 & m_2 & m_3 & m_4
\end{matrix}\right)^{(i)}\equiv\sum_{m=-i}^i(-1)^{i-m}\left(\begin{matrix}
j_1& j_2& i\\
m_1& m_2&m
\end{matrix}\right)
\left(\begin{matrix}
i& j_3& j_4\\
-m& m_3&m_4
\end{matrix}\right) \ .
\end{equation}
The states
\begin{equation}
\ket{i}=\sum_{m_1,m_2,m_3,m_4}\sqrt{2i+1}\left(\begin{matrix}
j_1 & j_2 & j_3 & j_4\\
m_1 & m_2 & m_3 & m_4
\end{matrix}\right)^{(i)}
\ket{j_1,m_1,j_2,m_2,j_3,m_3,j_4,m_4}
\end{equation}
are the orthogonal invariant states in the tensor product of the four representations $j_i$. They obey the orthogonality relations:
\begin{equation}\label{4jm_Cont}
\sum_{m_1,m_2,m_3,m_4}
\left(\begin{matrix}
j_1 & j_2 & j_3 & j_4\\
m_1 & m_2 & m_3 & m_4
\end{matrix}\right)^{(i)}
\left(\begin{matrix}
j_1 & j_2 & j_3 & j_4\\
m_1 & m_2 & m_3 & m_4
\end{matrix}\right)^{(i')}=\frac{\delta_{ii'}}{d_i} \ .
\end{equation}
In this work we adopted the compact notation \cite{Speziale:2016axj}:
\begin{equation}
\left(\begin{matrix}
j_i \\
m_i
\end{matrix}\right)=
\left(\begin{matrix}
j_i \\
m_i
\end{matrix}\right)^{(i)}=
\left(\begin{matrix}
j_1 & j_2 & j_3 & j_4\\
m_1 & m_2 & m_3 & m_4
\end{matrix}\right)^{(i)} \ .
\end{equation}
The integration over the Haar measure of four Wigner matrices can be expressed in term of $4jm$-symbols as:
\begin{equation}
\int_{SU(2)}duD^{j_1}_{m_1n_1}(u)D^{j_2}_{m_2n_2}(u)D^{j_3}_{m_3n_3}(u)D^{j_4}_{m_4n_4}(u)=\sum_j(2j+1)\left(\begin{matrix}
j_i \\
m_i
\end{matrix}\right)^{(i)}\left(\begin{matrix}
j_i \\
n_i
\end{matrix}\right)^{(i)} \ .
\end{equation}
The vertex amplitude includes the ${15j}$-symbols of the first kind \cite{Yutsis:1962vcy}, which can be expressed as the contraction over their magnetic indices of the product of five $4jm$-symbols:
\begin{equation}\label{15j}
\begin{split}
\{15j\}(i_k,j_a)\equiv\sum_{m_a}&
\left(\begin{matrix}
j_1 & j_2 & j_3 & j_4\\
m_1 & m_2 & m_3 & m_4
\end{matrix}\right)^{(i_1)}
\left(\begin{matrix}
j_1 & j_5 & j_6 & j_7\\
m_1 & m_5 & m_6 & m_7
\end{matrix}\right)^{(i_2)}
\left(\begin{matrix}
j_7 & j_2 & j_8 & j_9\\
m_7 & m_2 & m_8 & m_9
\end{matrix}\right)^{(i_3)}\\
&\left(\begin{matrix}
j_9 & j_6 & j_3 & j_{10}\\
m_9 & m_6 & m_3 & m_{10}
\end{matrix}\right)^{(i_4)}
\left(\begin{matrix}
j_{10} & j_8 & j_5 & j_4\\
m_{10} & m_8 & m_5 & m_4
\end{matrix}\right)^{(i_5)}
\end{split} \ .
\end{equation}

\section{Explicit proof of \eqref{eq:dSL2C_CG}}
\label{app:proof}


The starting point is the expression for the \sltc{} matrix elements \eqref{eq:dSL2C}. For better bookkeeping we will denote $z=e^{-r}$. We use the properties of the ${}_2F_1$ function to write it as the sum of two ${}_2F_1$ functions evaluated at $z^{-2}$, obtaining:
\begin{equation}
\begin{split}
d_{jlm}^{(\rho,k)}(r)&=(-1)^{j-l}\sqrt{\frac{\left(i\rho-j-1\right)!\left(j+i\rho\right)!}{\left(i\rho-l-1\right)!\left(l+i\rho\right)!}}\frac{\sqrt{(2j+1)(2l+1)}}{(j+l+1)!}z^{-(i\rho-k-m-1)}\\
&\sqrt{(j+k)!(j-k)!(j+m)!(j-m)!(l+k)!(l-k)!(l+m)!(l-m)!}\\
&\sum_{s,t}(-1)^{s+t}z^{2t}\frac{(k+s+m+t)!(j+l-k-m-s-t)!}{t!s!(j-k-s)!(j-m-s)!(k+m+s)!(l-k-t)!(l-m-t)!(k+m+t)!}\\
&\left\{\frac{(j+l+1)!(l-m-i\rho-k-s-t-1)!}{(l-i\rho)!(j+l-m-k-s-t)!}z^{-2(k+m+s+t+1)}\right.\\
&\ _{2}F_{1}\left[\{j+i\rho+1,k+m+s+t+1\},\{m+i\rho+k+s+t-l+1\};z^{-2}\right]\\
&+\frac{(j+l+1)!(m+i\rho+k+s+t-l-1)!}{(j+i\rho)!(k+m+s+t)!}z^{-2(l-i\rho+1)}\\
&\ _{2}F_{1}\left[\{j+l-m-k-s-t+1,l-i\rho+1\},\{l-m-i\rho-k-s-t+1\};z^{-2}\right]\bigg\} \ .
\end{split}
\end{equation}
We can write the two ${}_2F_1$ functions explicitly, and using the properties of the Pochhammer symbols, we obtain:
\begin{equation}
\begin{split}
d_{jlm}^{(\rho,k)}(z)&=(-1)^{j-l}\sqrt{(2j+1)(2l+1)}\sqrt{\frac{\left(i\rho-j-1\right)!}{\left(l+i\rho\right)!\left(j+i\rho\right)!}}\\
&\sqrt{(j+k)!(j-k)!(j+m)!(j-m)!(l+k)!(l-k)!(l+m)!(l-m)!}\\
&\Bigg\{ \frac{1}{(l-i\rho)!\sqrt{(i\rho-l-1)!}}\sum_{s,t,n}(-1)^{s+t}(-1)^{n}z^{-(i\rho+k+m+2s+2n+1)}\\
&\frac{(j+i\rho+n)!(k+m+s+t+n)!(l-m-i\rho-k-s-t-1-n)!}{t!s!n!(j-k-s)!(j-m-s)!(k+m+s)!(l-k-t)!(l-m-t)!(k+m+t)!}\\
&+\sqrt{(i\rho-l-1)!}\sum_{s,t,n}(-1)^{s+t}z^{2t+i\rho+k+m-2l-2n-1}\\
&\frac{(j+l-m-k-s-t+n)!(m+i\rho+k+s+t-l-1-n)!}{t!s!n!(j-k-s)!(j-m-s)!(k+m+s)!(l-k-t)!(l-m-t)!(k+m+t)!(i\rho-l-1-n)!}\Bigg\} \ .
\end{split}
\end{equation}
The summations over $s$ and $t$ can be decoupled by shifting the index $n\rightarrow n-s-k-m$ in the first sum and $n\rightarrow n+t+k-l$ in the second. Notice that this change of variable is well defined since $s$, $k+m$ and $k-l$ are all integers. Moreover, since the expression is getting quite lengthy we split it in two pieces
\begin{equation}
\label{eq:dsmallF1F2}
d_{jlm}^{(\rho,k)}(z)=\mathcal{F}_1+\mathcal{F}_2
\end{equation}
Where we defined 
\begin{equation}\label{eq:F1_Def}
\begin{split}
\mathcal{F}_{1}&=(-1)^{j-l}\sqrt{(2j+1)(2l+1)}\sqrt{\frac{(i\rho-l-1)!\left(i\rho-j-1\right)!}{\left(l+i\rho\right)!\left(j+i\rho\right)!}}\\
&\sqrt{(j+k)!(j-k)!(j+m)!(j-m)!(l+k)!(l-k)!(l+m)!(l-m)!}\\
&\sum_{s,t,n}(-1)^{s+t}z^{i\rho-k-1+m-2n}\\
&\frac{(j-m+n-s)!(i\rho+m-n-1+s)!}{s!(j-k-s)!(j-m-s)!(k+m+s)!}\\
&\frac{1}{t!(l-k-t)!(l-m-t)!(k+m+t)!(i\rho-k-1-n-t)!(k-l+n+t)!} \ ,
\end{split}
\end{equation}
and 
\begin{equation}\label{eq:F2_Def}
\begin{split}
\mathcal{F}_{2}&=(-1)^{j-l}\frac{\sqrt{(2j+1)(2l+1)}}{(l-i\rho)!}\sqrt{\frac{\left(i\rho-j-1\right)!}{\left(l+i\rho\right)!\left(j+i\rho\right)!(i\rho-l-1)!}}\\
&\sqrt{(j+k)!(j-k)!(j+m)!(j-m)!(l+k)!(l-k)!(l+m)!(l-m)!}\\
&\sum_{s,t,n}(-1)^{t+n-m-k}z^{-(i\rho-k+1-m+2n)}\\
&\frac{(i\rho-k+j+n-m-s)!}{s!(j-k-s)!(j-m-s)!(k+m+s)!(-k+n-m-s)!}\\
&\frac{(n+t)!(-i\rho-1-n-t+l)!}{t!(l-k-t)!(l-m-t)!(k+m+t)!} \ .
\end{split}
\end{equation}

We focus on $\mathcal{F}_{1}$ first. We shift one of the sum by $t\rightarrow t+l-k-n$ and we rearrange the terms:
\begin{equation}\label{eq:F1_Prima}
\begin{split}
\mathcal{F}_{1}&=\sum_{s,t,n}z^{i\rho-k-1+m-2n}\\
&\sqrt{\frac{(2j+1)(i\rho-j-1)!(j+k)!(j-k)!(j+m)!(j-m)!}{(j+i\rho)!}}\\
&(-1)^{j-k-n}(-1)^{s}\frac{(j-m+n-s)!(i\rho+m-n-1+s)!}{s!(j-k-s)!(j-m-s)!(k+m+s)!}\\
&\sqrt{\frac{(2l+1)(i\rho-l-1)!(l+k)!(l-k)!(l+m)!(l-m)!}{(l+i\rho)!}}\\
&(-1)^{t}\frac{1}{t!(i\rho-l-1-t)!(k-m+n-t)!(n-t)!(l+m-n+t)!(l-k-n+t)!}
\end{split}
\end{equation}
The $SU(2)$ Clebsch-Gordan coefficients can be expressed, using the Van der Waerden's formula \eqref{VanDerWa} and \cite{Yutsis:1962vcy}, in terms of the ${}_3 F_2$ hypergeometric function evaluated in 1
\begin{equation}
\label{eq:ComplexCG}
\begin{split}
\left\langle \left.j_1,m_1,j_2,m_2\right| j,m\right\rangle = & \sqrt{2j+1}\sqrt{\frac{(j_{1}+j_{2}-j)!(j_{1}-j_{2}+j)!(-j_{1}+j_{2}+j)!}{(j_{1}+j_{2}+j+1)!}}\\
&\sqrt{(j_{1}+m_{1})!(j_{1}-m_{1})!(j_{2}+m_{2})!(j_{2}-m_{2})!(j+m)!(j-m)!}\\
& \sum_{t}(-1)^{t}\frac{1}{t!(j_{1}+j_{2}-j-t)!(j_{1}-m_{1}-t)!(j_{2}+m_{2}-t)!(j-j_{2}+m_{1}+t)!(j-j_{1}-m_{2}+t)!} \\
=& \sqrt{2j+1}\sqrt{\frac{(j_{1}-j_{2}+j)!(-j_{1}+j_{2}+j)!}{(j_{1}+j_{2}+j+1)!(j_{1}+j_{2}-j)!}}\\
& \sqrt{\frac{(j_{1}+m_{1})!(j_{2}-m_{2})!(j+m)!(j-m)!}{(j_{1}-m_{1})!(j_{2}+m_{2})!}}\frac{1}{(j-j_{1}-m_{2})!(j-j_{2}+m_{1})!}\\
&{}_{3}F_{2}(j-j_{1}-j_{2},m_{1}-j_{1},-j_{2}-m_{2};j-j_{2}+m_{1}+1,j-j_{1}-m_{2}+1;1) \ .
\end{split}
\end{equation}
The ${}_3 F_2$ admits a well-defined analytic continuation for complex parameters, and as a consequence a well-defined analytic continuation of the Clebsch-Gordan coefficients.

We can perform the sums over $s$ and $t$ in \eqref{eq:F1_Prima} exactly in terms of ${}_3 F_2$ hypergeometric functions evaluated in 1. Using an identity of the ${}_3 F_2$ functions and manipulating the factorials in front we can recognize the two sums as analytically continued Clebsch-Gordan coefficients\footnote{The precise manipulation are quite cumbersome to report here. We refer to the Master thesis of one of the autors \cite{TesiAlessandro} for a step by step description. Note that the names $\mathcal{F}_1$ and $\mathcal{F}_2$ have been reversed.}
\begin{equation}\label{F1_Final}
\begin{split}
\mathcal{F}_{1}&=\sum_{n}z^{i\rho-k-1+m-2n}\\
&\left\langle\left.\left(\frac{i\rho+k-1}{2},m-n+\frac{i\rho-k-1}{2}\right),\left(\frac{i\rho-k-1}{2},n-\frac{i\rho-k-1}{2}\right)\right|j,m\right\rangle\\ 
&\left\langle\left.\left(\frac{i\rho+k-1}{2},m-n+\frac{i\rho-k-1}{2}\right),\left(\frac{i\rho-k-1}{2},n-\frac{i\rho-k-1}{2}\right)\right|l,m\right\rangle \ .
\end{split}
\end{equation}

Similarly, in $\mathcal{F}_2$ we can perform the sums over $s$ and $t$  in terms of two ${}_3F_2$ functions evaluated in $1$ and we can identify the analytically continued Clebsch-Gordan coefficients
\begin{equation}
\begin{split}
\mathcal{F}_{2}&=\sum_{n}z^{-(i\rho-k+1-m+2n)}\\
&\left\langle\left.\left(\frac{-i\rho-k-1}{2},\frac{-i\rho+k-1}{2}+m-n\right),\left(\frac{-i\rho+k-1}{2},n-\frac{-i\rho+k-1}{2}\right)\right|j,m\right\rangle\\ 
&\left\langle\left.\left(\frac{-i\rho-k-1}{2},\frac{-i\rho+k-1}{2}+m-n\right),\left(\frac{-i\rho+k-1}{2},n-\frac{-i\rho+k-1}{2}\right)\right|l,m\right\rangle
\end{split}
\end{equation}
Inserting the expression of $\mathcal{F}_1$ and $\mathcal{F}_2$ in \eqref{eq:dsmallF1F2} we obtain an expression of the reduced matrix elements in the $(\rho,k)$ representation in terms of complex Clebsch-Gordan coefficients
\begin{equation}\label{eq:SmallD_SL2C_Final}
\begin{split}
d_{jlm}^{(\rho,k)}(r)=&\sum_{n}e^{-(i\rho-k-1+m-2n)r}\\
&\left\langle\left.\left(\frac{i\rho+k-1}{2},\frac{i\rho-k-1}{2}+m-n\right),\left(\frac{i\rho-k-1}{2},n-\frac{i\rho-k-1}{2}\right)\right|j,m\right\rangle\\
&\left\langle\left.\left(\frac{i\rho+k-1}{2},\frac{i\rho-k-1}{2}+m-n\right),\left(\frac{i\rho-k-1}{2},n-\frac{i\rho-k-1}{2}\right)\right|l,m\right\rangle \\
+&\sum_{n}e^{-(-i\rho+k-1+m-2n)r}\\
&\left\langle\left.\left(\frac{-i\rho-k-1}{2},\frac{-i\rho+k-1}{2}+m-n\right),\left(\frac{-i\rho+k-1}{2},n-\frac{-i\rho+k-1}{2}\right)\right|j,m\right\rangle\\ 
&\left\langle\left.\left(\frac{-i\rho-k-1}{2},\frac{-i\rho+k-1}{2}+m-n\right),\left(\frac{-i\rho+k-1}{2},n-\frac{-i\rho+k-1}{2}\right)\right|l,m\right\rangle \ .
\end{split}
\end{equation}

\end{appendices}


\printbibliography

\end{document}